\documentclass[twocolumn,prd,superscriptaddress,showpacs,floatfix,%
preprintnumbers, nofootinbib]{revtex4}  
\usepackage{epsfig}   
\usepackage{bm}  
 \usepackage{multirow}  
\usepackage{color}

\newcommand{\be}{\begin{eqnarray}}  
\newcommand{\ee}{\end{eqnarray}}

\def\ltap{\ \raisebox{-.4ex}{\rlap{$\sim$}} \raisebox{.4ex}{$<$}\ }  

\def\lsim{{\ltap}}

\begin{document}  

\title{Signatures of collective and matter effects on
supernova neutrinos at large detectors}
  
\author{Sandhya Choubey}   
\affiliation{Harish-Chandra Research Institute, 
Chhatnag Road, Jhunsi, Allahabad  211019, India}  
   
\author{Basudeb Dasgupta}   
\affiliation{CCAPP, The Ohio State University, 191 W. Woodruff Avenue, Columbus 43210 Ohio, U.S.A }    
   
\author{Amol Dighe}   
\affiliation{Tata Institute of Fundamental Research, 
Homi Bhabha Road, Mumbai 400005, India}   
   
\author{Alessandro Mirizzi}   
\affiliation{II. Institut f\"ur Theoretische Physik, 
Universit\"at Hamburg, Luruper Chaussee 149, 
22761 Hamburg, Germany}
   
\date{August 1, 2010}   
   
\preprint{TIFR/TH/10-20}

\begin{abstract}   
We calculate the expected galactic supernova neutrino signal at 
large next-generation underground detectors. At different epochs after the 
explosion, the primary fluxes can be quite different. For these primary neutrino 
fluxes, spectral splits induced by collective neutrino flavor transformations can arise for either mass hierarchy in both neutrino and antineutrino 
channels. We classify flux models 
according to the nature and number of these splits, and calculate the 
observable $\nu_e$ and $\bar{\nu}_e$ spectra at Earth, taking into account subsequent matter effects. We find that 
some of the spectral splits could occur sufficiently close to the peak energies to produce 
significant distortions in the observable SN neutrino signal.  The most striking 
signature of this effect would be presence of  peculiar energy dependent modulations associated with Earth matter
crossing, 
present only in portions of the SN neutrino energy spectra demarcated by spectral 
splits. These signatures at proposed large water Cherenkov, scintillation, and liquid 
Argon detectors could give hints about the primary SN neutrino fluxes, as well as on the 
neutrino mass hierarchy and the mixing angle~$\theta_{13}$.

\end{abstract}   
   
   
\maketitle

\section{Introduction}         \label{intro}  

Observable effects of supernova (SN) neutrinos\footnote{Neutrinos will refer to both neutrinos ($\nu$) and antineutrinos ($\bar{\nu}$) in this article.} in underground detectors  
are a subject of intense investigation in astroparticle physics. In 
particular, a lot of attention has been devoted to possible signatures  
of the Mikheyev-Smirnov-Wolfenstein (MSW) matter effect~\cite{Matt} on 
neutrino flavor evolution~\cite{Raffelt:2010zz, Dighe:2009nr,Duan:2009cd}. 
Shock-wave effects in the stellar envelope~\cite{Taka, Luna, Foglish, 
Tomas, Dasgupta, FogliMega, Barger:2005it, Fogli:2006xy, Choubey:2006aq, 
Choubey:2007ga, Friedland:2006ta, Kneller:2007kg,Kneller:2010sc} and the neutrino Earth matter 
crossing~\cite{Dighe, Lunardini:2001pb, Takahashi:2001dc, Dighe:2003jg, 
Dighe:2003vm, Lunardini:2003eh, Mirizzi:2006xx} have been predicted to 
have dramatic effects on SN neutrino oscillations and to produce 
signatures that could allow us to extract valuable information on the 
unknown neutrino mass and mixing parameters, like the neutrino mass
hierarchy and the mixing angle~$\theta_{13}$.   

In the recent years, it has been understood that the paradigm of neutrino flavor 
transformations in supernovae, based on the only MSW effect with the 
ordinary matter~\cite{Dighe}, was incomplete. Novel phenomena have been 
found to be important in the region close to the neutrinosphere, 
where the neutrino density is so high that effects of neutrino-neutrino 
interactions dominate the flavor evolution. These effects have been 
understood and characterized in an increasingly realistic way in a long series of 
papers~\cite{Pantaleone:1992eq, Samuel:1993uw, Qian:1995ua, 
Pastor:2002we, Sawyer:2005jk, Fuller:2005ae, Duan:2005cp, Duan:2006an, 
Hannestad:2006nj, Duan:2007mv, Raffelt:2007yz, EstebanPretel:2007ec, 
Raffelt:2007cb, Raffelt:2007xt, Duan:2007fw, Fogli:2007bk, Duan:2007bt, 
Duan:2007sh, Dasgupta:2008cd, EstebanPretel:2007yq, Dasgupta:2007ws, 
Duan:2008za, Dasgupta:2008my,Sawyer:2008zs, Duan:2008eb, Chakraborty:2008zp, 
Lunardini:2007vn, 
Dasgupta:2008cu, EstebanPretel:2008ni,Gava:2008rp, Fogli:2008pt, 
Duan:2008fd, Raffelt:2008hr, Blennow:2008er, Fogli:2008fj, Gava:2009pj, 
Dasgupta:2009mg, Fogli:2009rd, Galais:2009wi, Mirizzi:2009td, 
EstebanPretel:2009is, Chakraborty:2009ej,Duan:2010bg,Dasgupta:2010ae,Friedland:2010sc,
Dasgupta:2010cd,Duan:2010bf,Cherry:2010yc}.
   
The neutrino-neutrino interactions provide a large potential   
for the neutrinos due to the neutrinos themselves, which causes large and 
rapid conversions between different flavors. The transitions occur collectively, 
i.e. in a coherent fashion, over the 
entire energy range. 
Unlike ordinary neutrino oscillations, these collective oscillations depend 
strongly on the original SN neutrino fluxes. In a supernova $\nu_e$ and $\bar{\nu}_e$ are  distinguished from other flavors due to their additional charged-current interactions. The $\nu_{\mu,\tau}$ and their antiparticles,
on the other hand, are produced at practically identical rates.
Following the standard terminology, we define the two relevant non-electron flavor states as $\nu_{x,y}=\cos\theta_{23}\nu_\mu\mp\sin\theta_{23}\nu_\tau$, where $\theta_{23} \simeq \pi/4$ is the
atmospheric mixing angle. Since the initial $\nu_x$ and
$\nu_y$ fluxes are identical, 
the primary neutrino 
fluxes are best expressed in terms of $\nu_e$, $\bar{\nu}_e$ and $\nu_x$.
The collective pair conversions 
$\nu_e {\bar \nu}_e  \leftrightarrow \nu_{x,y} {\bar \nu}_{x,y}$
take place even for extremely small mixing angles
within the first ${\cal O}(500)$~km~\cite{Hannestad:2006nj},
much before the MSW flavor conversions start. 
In a typical supernova, enough asymmetry between 
neutrino and antineutrino number fluxes is expected, 
thus complicated angle-dependent 
decoherence  effects due to the anisotropic neutrino emission in the 
SN environment are likely to be suppressed~\cite{EstebanPretel:2007ec}.   
There may be however a slow-down of  collective flavor conversions  due to 
multi-angle effects in deep supernova regions~\cite{Duan:2010bf}.

During the accretion phase ($t \lesssim 0.5 $~ms after the core-bounce) in typical supernova simulations~\cite{livermore, garching, Keil:2002in, Raffelt:2003en, Fischer:2009af} 
one finds  an almost perfectly equipartitioned luminosity between 
the neutrino flavors:
$L_{\nu_e} \approx L_{{\bar \nu}_e} \approx L_{{\nu}_x}$.
This has been the  benchmark scenario 
in most studies of collective flavor oscillations, which predicts: 
For normal mass hierarchy (NH) $\nu$ and $\bar\nu$ remain unaffected by 
collective oscillations. For Inverted hierarchy (IH), the end of 
collective oscillations is marked  by an almost complete exchange of the 
$\bar{\nu}_e$ and $\bar{\nu}_y$ flavor spectra for ${\bar\nu}$'s, while for $\nu$'s 
the  swap occurs only above a characteristic energy fixed  by lepton number conservation, 
giving rise to a spectral~split in the  $\nu$ energy 
distributions~\cite{Raffelt:2007cb,Duan:2007fw,Duan:2008fd}.

It is not obvious that the neutrino fluxes maintain { their approximate equipartition even} at late times. Indeed, the deleptonization of the core is probably faster than the cooling of the proto-neutron star, so that the asymmetry between 
the fluxes can become smaller and the spectra at late times depend 
on details of the neutrino transport (the density and the temperature 
profiles, as well as the treatment of the interaction rates). Due to this 
complexity, results from  SN simulations are not unambiguous.  
Late-time cooling calculations  performed by the 
Garching group~\cite{Keil:2002in}  show a cross-over of the different 
neutrino luminosities $L_{\nu_x}\gtrsim L_{{\nu}_e} \approx L_{{\bar \nu}_e}$, 
i.e. at late times the $\nu_x$ flux becomes relatively larger~\cite{Raffelt:2003en}. 
Similar results, but less pronounced, have also been obtained 
recently by the Basel group~\cite{Fischer:2009af}
and in the most recent long-time simulations of the Garching group 
\cite{Huedepohl:2009wh}.

This latter case ($L_{\nu_x}\gtrsim L_{{\nu}_e} \approx L_{{\bar \nu}_e}$) was recently studied~\cite{Dasgupta:2009mg,Friedland:2010sc,
Dasgupta:2010cd}, and one found
the occurrence of unexpected multiple spectral swaps and consequent 
spectral splits for both $\nu$'s and ${\bar\nu}$'s in either mass hierarchy.
The lesson from this result is that the benchmark scenario (one spectral split 
in $\nu_e$ spectrum and complete swap in the ${\bar \nu}_e$ for inverted 
hierarchy) is in fact a special case, while the phenomenology of the spectral 
features can be more complex. To this end, a detailed study has been performed 
in~\cite{Fogli:2009rd}, where the impact of the variations  of the neutrino 
luminosities has been explored, finding abrupt changes in the number and the 
position of  the splits by slightly changing the ratio between the luminosities
of the different neutrino species around some critical value. 

Given the sensitivity of the self-induced spectral splits on the original 
SN neutrino emission features, their detection in a galactic SN neutrino 
burst could provide a tool to reconstruct the original neutrino fluxes. Since 
spectral splits could potentially affect both neutrinos and antineutrinos, 
it seems worthwhile to investigate complementary detection techniques with 
sensitivity to  $\nu_e$ and ${\bar \nu}_e$ respectively.
Around the world, there is an active discussion about the 
feasibility of three different classes of large underground detectors for 
low-energy neutrino physics and astrophysics, \emph{viz.} water Cherenkov 
detectors with fiducial masses ranging from  
${\mathcal O}(0.4\,\ \textrm{Mt})$~\cite{Jung:1999jq, 
Nakamura:2003hk, deBellefon:2006vq} to 5~Mt~\cite{Kistler:2008us}, liquid 
scintillation detectors with masses of 
${\mathcal O}(50\,\ \textrm{kt})$~\cite{MarrodanUndagoitia:2008zz} and 
liquid Argon Time Projection Chambers (LAr TPC) with fiducial masses
of ${\mathcal O}(100\,\  \textrm{kt})$~
\cite{Cline:2006st, Badertscher:2008bp, 
Ereditato:2005ru, Baibussinov:2007ea}. 
In particular, these three detection techniques are the backbones of 
the European project LAGUNA (Large Apparati 
for Grand Unification and Neutrino Astrophysics)~\cite{Autiero:2007zj}
and the LBNE (Long Baseline Neutrino Experiment) in 
DUSEL (Deep Underground Science and Technology Laboratory) 
~\cite{lbne-dusel}. 
The physics potential of such devices for supernova neutrino detection would
be extremely high. In particular, water Cherenkov and scintillation 
neutrino experiments  are mostly sensitive to supernova $\bar{\nu}$
through the inverse beta decay process ${\bar \nu}_e + p \to  e^{+} + n$. 
On the other hand, LAr TPC would have a high sensitivity to SN $\nu$, 
through the charged current interactions of $\nu_e$ with the Ar nuclei in 
the detector.  The complementarity between these different detection
techniques would allow to compare the different features, like the spectral
splits, occurring in SN neutrino spectra.

Motivated by this intriguing perspective, our aim in this paper is to calculate 
the observable SN signal at these detectors. The main feature of our work is an improved treatment of the oscillation physics, particularly collective effects, in different phases of the SN explosion. We also investigate the detectability of possible features in the observable galactic supernova signal at  future large underground detectors  and examine the 
physics potential.

The plan of the article is as follows. 
In Sec.~\ref{sec:initial_conditions}, we describe the 
parameter space of expected primary neutrino fluxes, 
the equations for the neutrino flavor evolution, and neutrino mixing.
In Sec.~\ref{sec:collective}, 
we give a short overview of the collective effects on neutrino
spectra and the multiple spectral splits. 
We also perform a scan on the SN flux parameter space to identify 
different classes of fluxes that lead to qualitatively different final spectra.
Sec.~\ref{sec:afterMSW} discusses the subsequent flavor conversions
due to MSW effects and possible Earth matter effects, and
presents the net survival probabilities for $\nu_e$ and $\bar\nu_e$
arriving at the detectors.
 In Sec.~\ref{sec:detectors}, we describe the features of our reference 
detectors (fiducial mass, cross sections and energy resolution). 
In Sec.~\ref{sec:events} we present our results on the detectability 
of the spectral signatures in the $\nu_e$ and ${\bar \nu}_e$ signal. 
In particular, we discuss the distinctive features due to interplay of 
Earth matter effects and spectral splits. We also briefly comment on 
neutrino oscillation effects on the early neutronization burst and 
in relation to the SN shock-wave propagation.
Finally, in Sec.~\ref{sec:conclusions} we comment about our results,   
discuss future perspectives, and conclude.

\section{Fluxes, potentials and mixing parameters}
\label{sec:initial_conditions}

In this section, we describe the primary fluxes, the effective 
potentials they encounter during their propagation, and the neutrino
mixing parameters that we use in our numerical simulations.

\subsection{Primary neutrino fluxes}  
    
The primary neutrino fluxes of all neutrino and antineutrino
species can be written as 

\begin{equation}  
\label{Ybeta} F_\nu^0(E)= {\Phi^0_{\nu}}\,\varphi(E)\ ,  
\end{equation}  
where   $\nu = \nu_e, {\overline\nu}_e, \nu_x$, 
 $\varphi(E)$ is the normalized primary neutrino spectrum, i.e. $\int dE  
\; \varphi(E)=1$, and $\Phi_{\nu}^0$ is the total number 
flux.
For $\varphi(E)$ we use the spectral parameterization given in 
Ref.~\cite{Keil:2002in}:  
\begin{equation}  
\label{varphi} \varphi(E)=\frac{(\alpha_\nu+1)^{(\alpha_\nu+1)}}{  
\Gamma(\alpha_\nu+1)}\left(\frac{E}{\langle E_\nu  
\rangle}\right)^{\alpha_\nu} \frac{e^{-(\alpha_\nu+1)E/\langle  
E_\nu\rangle}}{\langle E_\nu \rangle}\ ,  
\end{equation}  
where $\langle E_\nu\rangle$ is the average $\nu$ energy, $\alpha_\nu$
 is a spectral parameter, and $\Gamma$ is the Euler gamma function. 
The values of the parameters are 
model dependent \cite{livermore,garching}. There is significant variation 
across different SN simulations, but we find that the 
$\langle E_{\nu_e} \rangle$ 
and $\langle E_{\bar\nu_e} \rangle$ are usually close to
\begin{equation}  
\label{Eave} \langle E_{\nu_e} \rangle=12~{\rm MeV~and~}\langle E_{\bar\nu_e}  
\rangle=15~{\rm MeV}\ ,  
\end{equation}  
whereas $\langle E_{\nu_x} \rangle$ tends to vary in the range
\begin{equation}
16 \ \mathrm{MeV} \leq \langle E_{\nu_x} \rangle\leq 25 \ \mathrm{MeV}\ .
\label{eq:ener}
\end{equation}
We fix the spectral pinching parameter at $\alpha_\nu=3$ for all
species.

The total number fluxes $\Phi_\nu^0$ are related to the luminosities
of the neutrino species through
\begin{equation}  
\Phi_{\nu}^0 = \frac{L_\nu}{\langle E_\nu \rangle} \,\ .
\end{equation}
For definiteness, we assume a typical supernova which emits
$E_B \approx  3\times10^{53}$~erg in a duration of $t=10$~s. 
The time integrated luminosities of all species combined is equal
to the total emitted energy:
\begin{equation}
\int dt (L_{\nu_e} + L_{{\bar\nu}_e} + 4 L_{{\nu}_x}) = E_B \;,
\end{equation}
such that the average number fluxes are about $10^{55}$~s$^{-1}$MeV$^{-1}$,
and about $10^{58}$ neutrinos are emitted in the entire SN explosion.

The ratio between the different luminosities varies significantly 
across models, but we find that 
(See Fig.3 of \cite{Fogli:2009rd} and Table 7.3 of~\cite{Keil:2003sw} 
for example)
\begin{eqnarray}
&L_{\nu_e} / L_{{\bar\nu}_e} \approx 1 \; ,&\nonumber \\
&0.5 \lesssim L_{\nu_x}/L_{\nu_e} \lesssim 2.0 \;.&
\label{eq:lumen}
\end{eqnarray}
The relative number fluxes change over the duration of the
neutrino burst. In particular, the fluxes in the accretion phase
are not necessarily similar to those in the cooling phase. 
These primary fluxes are subject to flavor conversions, of which 
the collective oscillations that we will discuss in the next section 
depend quite sensitively on the fluxes themselves.

\subsection{Neutrino equations of motion}
\label{sec:eom}

Mixed neutrinos are described by matrices of density
$\rho_{\bf p}$ and ${\bar \rho}_{\bf p}$ for each (anti)neutrino mode. 
The diagonal entries are the usual occupation numbers whereas 
the off-diagonal terms encode phase information. The equations
of motion (EoMs) are~\cite{Sigl:1992fn}
\begin{equation}
\textrm{i}\partial_t \rho_{\bf p} = [{\sf H}_{\bf p}, \rho_{\bf p}] \; ,
\end{equation}
where the Hamiltonian is 
\begin{equation}
{\sf H}_{\bf p} = {\sf \Omega}_{\bf p} + {\sf V}_{\rm MSW} + {\sf V}_{\nu\nu} \; .
\end{equation}
Here ${\sf \Omega}_{\bf p}$ is the matrix of the vacuum oscillation frequencies 
for neutrinos, which is
\be
{\sf \Omega}_{\bf p}= \textrm{diag}(m_1^2,m_2^2,m_3^2)/2|{\bf p}| \; 
\ee
in the mass basis.
For antineutrinos ${\sf \Omega}_{\bf p} \to -{\sf \Omega}_{\bf p}$.
The matter effect due to the background electron density $n_e$ is,
in the weak interaction basis, i.e. $(\nu_e,\nu_\mu,\nu_\tau)$, 
\be
{\sf V}_{\rm MSW}=\sqrt{2}G_F n_e \textrm{diag}(1,0,0) \; ,
\ee
neglecting the second-order difference between the $\nu_\mu$
and  $\nu_\tau$ refractive indices that could be important for
collective neutrino oscillations only at really early post-bounce times 
($t \lesssim 300$~ms) ~\cite{EstebanPretel:2007yq,Botella:1986wy}.

The effective potential due to the neutrino-neutrino interactions is
\be
{\sf V}_{\nu\nu} = \sqrt{2} G_F\int \frac{\textrm{d}^3 {\bf q}}{(2\pi)^3}
(\rho_{\bf q}-{\bar\rho}_{\bf q})
(1-{\bf v}_{\bf p}\cdot {\bf v}_{\bf q}) \,\ ,
\ee
where ${\bf v}_{\bf p}$ is the velocity of the neutrino mode ${\bf p}$.

In spherical symmetry the EoMs can be expressed as a closed set of 
differential equations along the radial direction
~\cite{EstebanPretel:2007ec,Dasgupta:2008cu}.
The factor $(1-{\bf v}_{\bf q}\cdot {\bf v}_{\bf q})$ in the Hamiltonian 
gives rise to ``multi-angle'' effects for neutrinos moving on different 
trajectories~\cite{Duan:2006an}. 
However, for realistic supernova conditions the modifications are presumably small 
(except for a shift in the onset of the flavor conversions~\cite{Duan:2010bf}), 
allowing for a single-angle approximation~\cite{Fogli:2008fj,EstebanPretel:2007ec}. 
We implement this approximation by assuming all neutrinos
to have been launched from the neutrinosphere at radius $R$,
with an angle $45^\circ$ relative to the 
radial direction~\cite{EstebanPretel:2007ec,EstebanPretel:2007yq}.
In this case, the formal derivation of the single-angle approximation
explicitly gives the radial dependence of the neutrino-neutrino
interaction strength as~\cite{Dasgupta:2008cu}
\begin{equation}
{\sf V}_{\nu\nu} (r) \propto\mu(r) = \mu_0 \frac{R^2}{r^2} C_r \,\ .
\end{equation}
Here, $\mu_0$ is the interaction strength 
\begin{equation}
\mu_0 = \sqrt{2}G_F (F_{\nu_e}^0-F_{\bar\nu_e}^0) \; ,
\end{equation}
the $r^{-2}$ scaling comes from the geometrical flux dilution,
while the collinearity factor
\begin{eqnarray}
C_r &=& 4 \left[\frac{1-\sqrt{1-(R/r)^2}}{(R/r)^2}\right]^2 -1  \nonumber \\
& = & \frac{1}{2}\left(\frac{R}{r}\right)^2 \,\ \,\ \,\ \,\ \textrm{for} \,
\  r\to\infty \,\ 
\end{eqnarray}
arises from the $(1- {\bf v}_{\bf p} \cdot {\bf v}_{\bf q})$ 
structure of the neutrino-neutrino interaction.
The decline of the neutrino-neutrino interaction strength, 
$\mu(r) \sim r^{-4}$ for $r\gg R$, is evident, while the numerical
coefficient for large $r$ depends on the launching angle,
here taken to be $45^\circ$.
When fluxes at the neutrinosphere radius $R=10$~km are taken,
we get 
\begin{equation}
\mu_0 = 7 \times 10^5 ~\textrm{km}^{-1} \,\ .
 \end{equation}
Of course, the physical
neutrinosphere is not a well-defined concept. Therefore,
the radius $R$ simply represents the location where
we fix the inner boundary conditions. However, essentially
nothing happens close to the neutrinosphere
because the in-medium mixing angle is extremely
small. Therefore, as far as the vacuum
and matter oscillation terms are concerned, it is almost
irrelevant where we fix the inner boundary condition.


\subsection{Neutrino mixing parameters}

We work in the rotated basis ~\cite{Dasgupta:2007ws} 
$$(\nu_e,\nu_x,\nu_y) \equiv {\sf R}^T(\theta_{23})
(\nu_e,\nu_\mu,\nu_\tau) \; .
$$
This is equivalent to taking $\theta_{23}=0$ in the neutrino mixing matrix,
which makes no difference to $\nu_e$ and $\bar\nu_e$ evolution as long as 
the primary fluxes of $\nu_\mu$ and $\nu_\tau$ are identical, { as we have assumed.} 
The values of the other mixing angles are $\sin^2\theta_{12} \simeq { 0.30}$ 
and $\sin^2 \theta_{13} \lesssim 0.04$~\cite{GonzalezGarcia:2010er}
in vacuum.
The neutrino mass-squared differences in vacuum are taken to be 
$\Delta m^2_{\rm atm}= 2\times 10^{-3}$~eV$^2$ and 
$\Delta m^2_{\rm sol}= { 7.6}\times 10^{-5}$~eV$^2$, 
close to their current best-fit values~\cite{GonzalezGarcia:2010er}. 
 We study both the cases of normal neutrino mass hierarchy
(NH: $\Delta m^2_{\rm atm} = m_3^2-m_{1,2}^2 >0$) 
and inverted mass hierarchy (IH: $\Delta m^2_{\rm atm} = m_3^2-m_{1,2}^2 <0$).
We ignore possible subleading CP violating effects~\cite{Gava:2008rp}
by setting $\delta_{\rm CP}=0$.

Matter effects in the region of collective oscillations 
(up to a few 100 km) suppress the mixing angles and slightly modify the neutrino
mass-square differences.
 We take the matter-suppressed mixing angles to be
$\tilde\theta_{13}=\tilde\theta_{12}= 10^{-3}$ and the effective mass-square differences 
$\Delta {\tilde m}^2_{\rm atm} = \Delta m^2_{\rm atm} \cos \theta_{13}\simeq \Delta m^2_{\rm atm}$
and $\Delta {\tilde m}^2_{\rm sol} = \Delta m^2_{\rm sol} \cos \theta_{12} \simeq 0.4 \Delta m^2_{\rm sol}$%
~\cite{Duan:2008za,EstebanPretel:2008ni}.
Apart from these shifts,  matter effects typically do not disturb the development of the collective neutrino
 oscillations,
except at very early times ($t\lesssim 300$~ms) when the effective electron density $n_e$ would become
larger than the neutrino density $n_\nu$,   suppressing the collective flavor conversions~\cite{EstebanPretel:2008ni}. 
For simplicity, in the following we will always focus on later times 
where collective oscillations are not inhibited by   a strong matter term.  MSW conversions   occur  after collective effects have 
ceased~\cite{Dasgupta:2007ws, Fogli:2008fj, Gava:2009pj}.
Therefore, their effects factorize, and will be described in
Sec.~\ref{sec:afterMSW}.

\begin{figure*}[!htpb]
\includegraphics[width=0.49\linewidth,clip=]{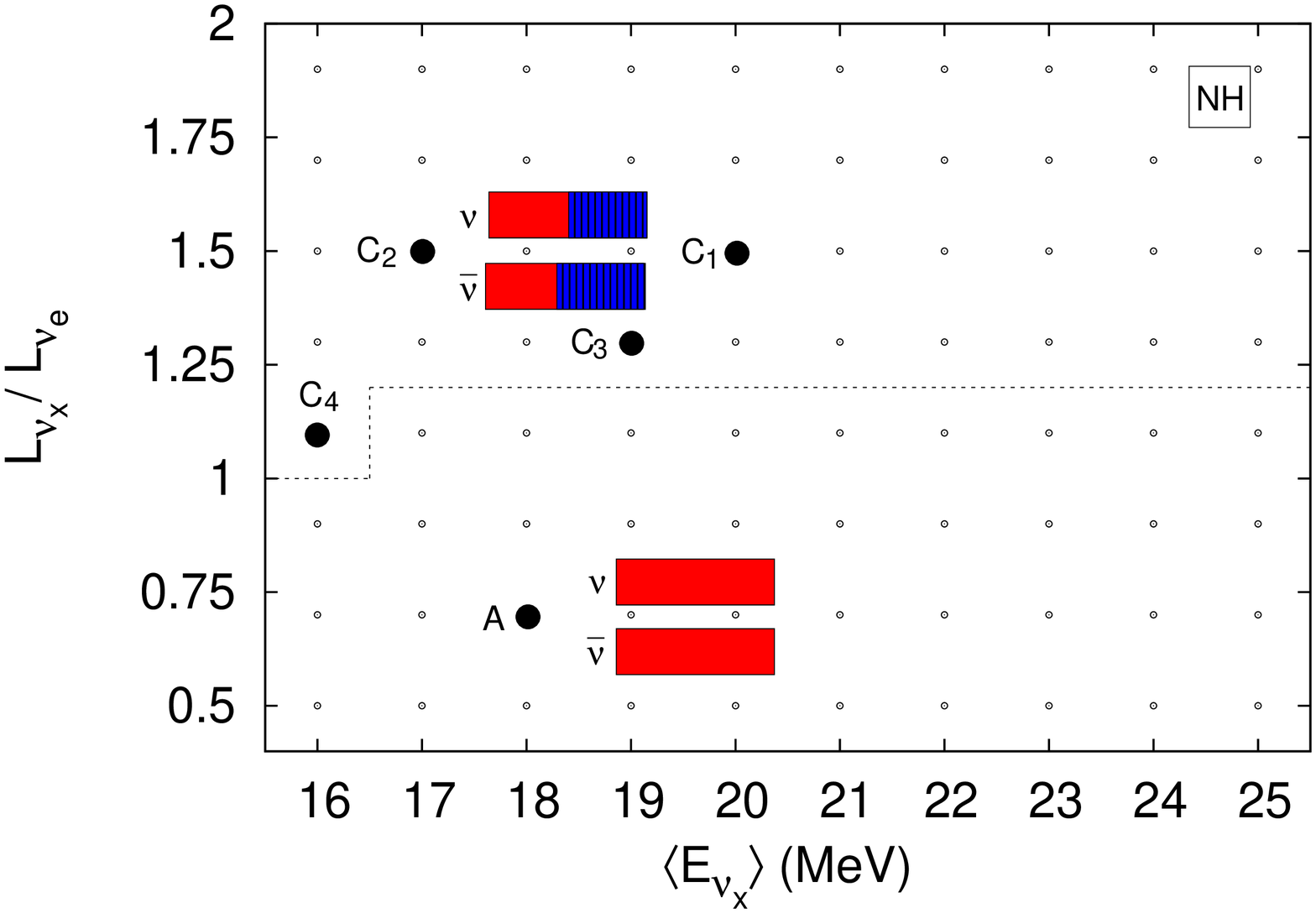}
\includegraphics[width=0.49\linewidth,clip=]{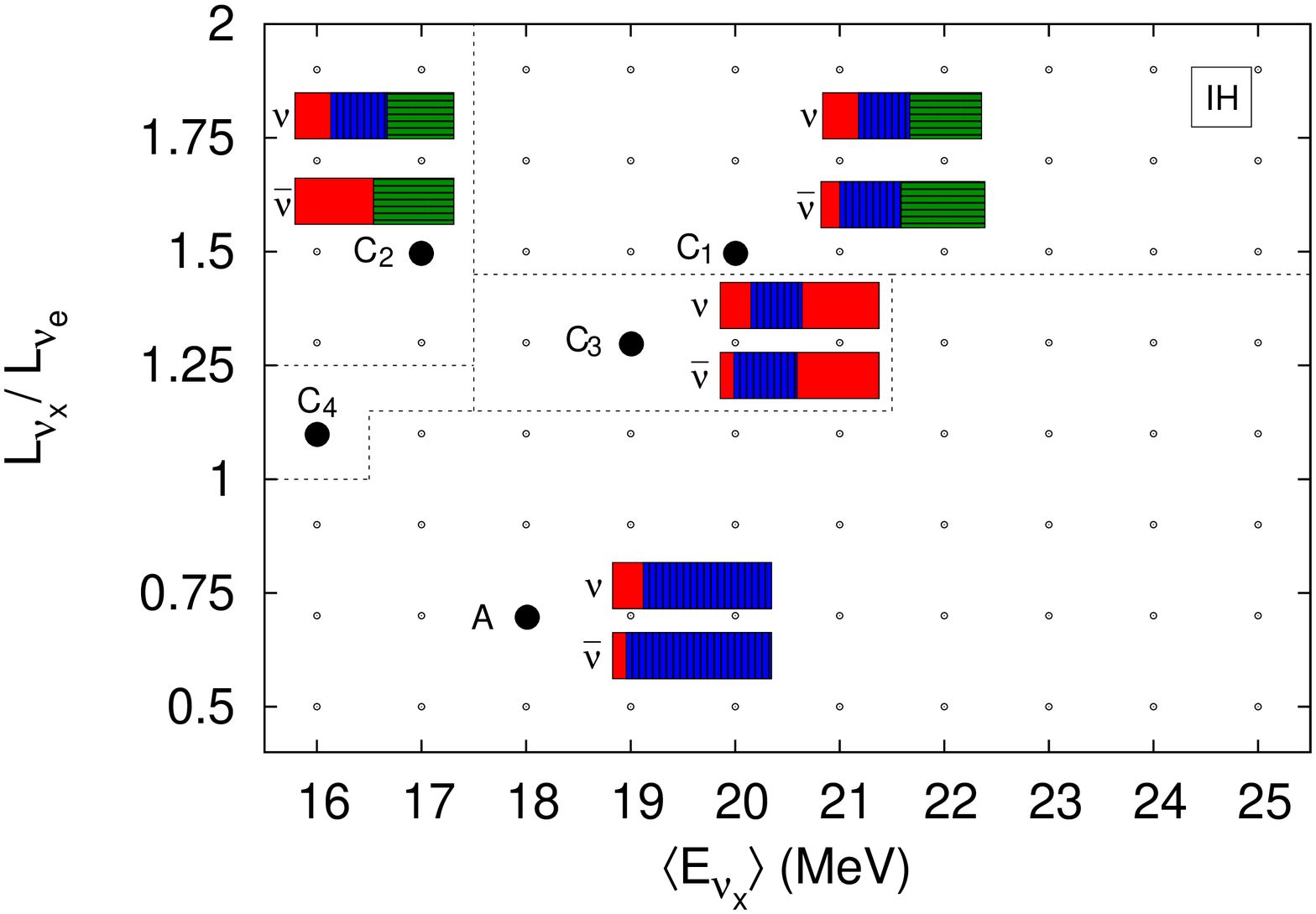}
\caption{Regions of the flux parameter space that produce qualitatively 
different spectra after collective flavor transformations. 
Left panel for NH, Right for IH. 
The bands represent the positions of spectral swaps in energy, and their
nature: vertical stripes correspond to $\nu_e \leftrightarrow \nu_y$ swap,
horizontal stripes correspond to $\nu_e \leftrightarrow \nu_x$ swap,
unstriped regions have no swap.
The widths of the striped / unstriped regions approximately indicate 
the energy range where they are present, the complete band
corresponding to the energy range 1-50 MeV.}
\label{fig:phasediag} 
\end{figure*}

\section{Collective flavor conversions}
\label{sec:collective}

Although the subject of collective oscillations has been discussed 
in great detail in number of papers, most have focused on the benchmark 
scenario used in~\cite{Fogli:2007bk}. 
It was pointed out in~\cite{Dasgupta:2009mg, Friedland:2010sc, Dasgupta:2010cd}
that the final outcome of collective effects can be quite complicated, 
depending on the primary fluxes.
In this section we will perform a detailed study investigating 
the effect of the primary fluxes on flavor oscillations
during the different post-bounce phases.

\subsection{Spectral swaps and  splits}

Near the neutrinosphere, due to the large neutrino density, the  
neutrino-neutrino interaction energy is very large. This ensures that   
neutrinos exhibit collective oscillations, i.e. neutrinos of all energies  
oscillate coherently with the same average frequency. 
As the neutrinos stream outward and the neutrino density becomes smaller, 
self-induced  oscillations begin to take place~\cite{Hannestad:2006nj}. 
{ These oscillations, being instability driven, have a large amplitude even  
for a very small mixing angle. 
In the presence of a slowly decreasing background neutrino density, 
the collective effects on the neutrinos become negligible beyond a point, and
their imprint on the spectra is left in the form of the spectral splits.
The energy and the number of the splits is crucially dependent on the 
neutrino mass hierarchy (normal or inverted), and on the relative sizes 
of the primary number fluxes of different flavors.}

The resultant dynamics of the spectral splits has been 
elaborated in~\cite{Dasgupta:2009mg,Dasgupta:2010cd}: 
spectral swaps can develop around \emph{unstable crossing points}.
Here crossing points are those energies $E_c$ 
where spectra of different flavors cross each other, i.e.
\begin{eqnarray}
F^0_{\nu_e}(E_c) = F^0_{\nu_x}(E_c) \quad \mbox{ or } \quad
F^0_{\bar\nu_e}(E_c) = F^0_{\nu_x}(E_c) \; .
\end{eqnarray}
A given crossing point is { unstable} for 
$e \leftrightarrow y$ (``atmospheric sector'') swap, triggered by 
$(\Delta m^2_{\rm atm},\theta_{13})$, if
at the energy $E_c$, 
\begin{eqnarray}
d(F^0_{\nu_e}-F^0_{\nu_x})/dE<0\quad{\rm for\;IH}\;, \nonumber \\
d(F^0_{\nu_e}-F^0_{\nu_x})/dE>0\quad{\rm for\;NH}\;,
\end{eqnarray}
and analogously for antineutrinos.
Thus, in IH we get a $\nu_e \leftrightarrow \nu_y$ or 
$\bar\nu_e \leftrightarrow \bar\nu_y$ swap around every crossing 
with a negative slope,
while in NH we get such a swap around every crossing with a positive slope.
Each swap is demarcated by two spectral splits. 
Two subtle points may be noted: 
(i) If two unstable crossings are very close to each other,
the swaps may be influenced by each other and even merge, and 
(ii) If an unstable crossing is too closely flanked by stable crossings, 
the width of the swap around it is suppressed exponentially, 
which also leads to adiabaticity violation leading to smearing out 
of that swap~\cite{Dasgupta:2009mg}.

In addition, a crossing point is unstable for the 
$e \leftrightarrow x$ (``solar sector'') swap, 
triggered by $(\Delta m^2_{\rm sol},\theta_{12})$, if~\cite{Dasgupta:2010cd}
\be
d(F^0_{\nu_e}-F^0_{\nu_x})/dE>0 \; ,
\ee
and analogously for antineutrinos.
We thus obtain a $\nu_e \leftrightarrow \nu_x$ or 
$\bar\nu_e \leftrightarrow \bar\nu_x$ swap around every crossing 
with a positive slope.
These swaps tend to be more non-adiabatic than those in
the atmospheric sector, since
the natural frequency of the solar sector, $\Delta m^2_{\rm sol}/(2E)$,
is smaller than that of the atmospheric sector.
As a result, the rate of change of the collective interaction
strength may become too fast for there to be enough time for
the solar sector instability to grow.

The combination of the atmospheric and solar sector swaps 
may give rise to three-flavor effects, pointed out recently
in~\cite{Friedland:2010sc,Dasgupta:2010cd}.
For NH, mass splittings are positive in both the
atmospheric and solar sectors, and both the mixing angles are
small due to matter suppression. 
Both the sectors then tend to produce instabilities at the
same energies, but the atmospheric sector wins due to its
higher natural frequency (and hence larger adiabaticity).
As a result, only $\nu_e \leftrightarrow \nu_y$ and
$\bar\nu_e \leftrightarrow \bar\nu_y$ swaps develop, and no
three-flavor effects are observed.
On the other hand, in IH the atmospheric and solar instabilities
are in different parts of the spectrum. 
The solar sector instability then operates in the high-energy region
without hindrance and causes an additional swap that may even 
merge with the atmospheric sector swap, partially erasing 
one of the spectral splits.   
However, the split usually is not completely erased because of the 
non-adiabaticity of the swap in the solar sector.


\begin{figure*}[!t]
\centering
\begin{tabular}{cc}
{\sf Flux Model A}& {\sf Flux Model C}$_1$\\
\epsfig{file=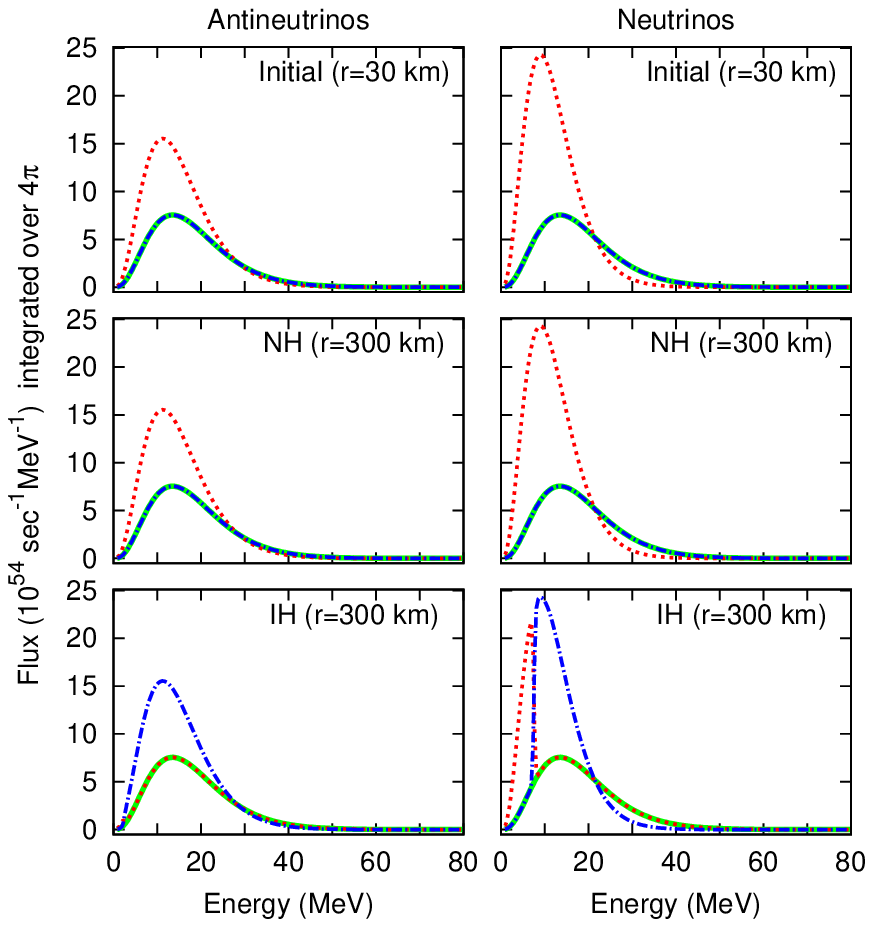,width=0.5\linewidth,clip=} &
\epsfig{file=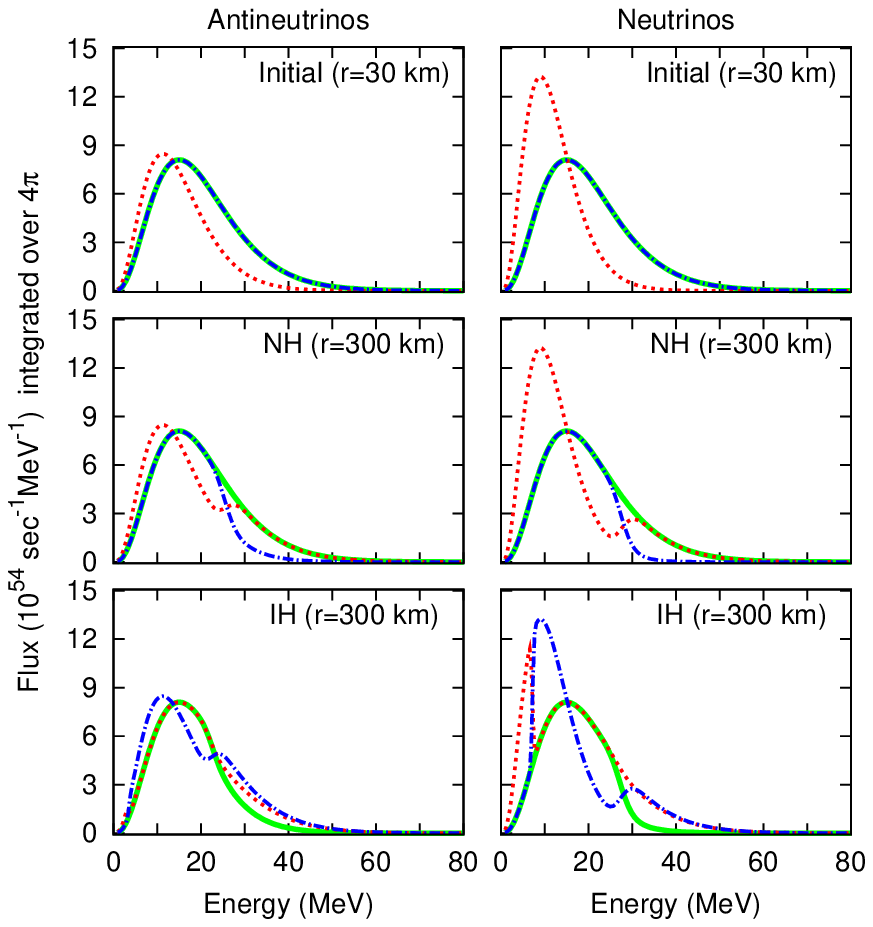,width=0.5\linewidth,clip=} \\
\end{tabular}
\caption{Fluxes (integrated over $4\pi$ solid angle) for the benchmark models: $\rm A$ (left panel), $\rm C_1$ (right panel). In each panel, fluxes are shown for antineutrinos (left) and neutrinos (right) at initial time at neutrinosphere (top),  post-collective oscillations in IH (middle) and NH (bottom). The $e$ flavor is shown in red/dotted, $x$ in green/solid, and $y$ in blue/dot-dashed. The fluxes for models $\rm C_2$, $\rm C_3$, $\rm C_4$ are not shown, as they are similar to $\rm C_1$, with certain features degraded (see text).}
\label{spectra_after_collective}
\end{figure*}

\subsection{Flux dependence of spectral splits}

The number and positions of the spectral splits depends on the primary
neutrino spectra~\cite{Fogli:2009rd}.
However a clear mapping between primary fluxes and
split patterns is still lacking.
The flux parameters with the widest variation among models
are the average energy of the $\nu_x$ flux $\langle E_{\nu_x}\rangle$,
and the $\nu_x$ luminosity $L_{\nu_x}$.
We scan the SN neutrino parameter space in the variables $L_{\nu_x}/L_{\nu_e}$ 
and $\langle E_{\nu_x} \rangle$ in the ranges defined by Eqs.~(\ref{eq:ener}) 
and (\ref{eq:lumen}), and numerically evolve the system of neutrinos to 
determine the regions { of the flux parameter space} 
where different spectral swaps are effective. 
We then classify the primary spectra depending on the number and nature
of these swaps. 

The results of this classification may be represented by a 
``phase diagram'' shown in Fig.~\ref{fig:phasediag}.
Note that the idea of a phase diagram, in which the oscillation patterns change 
abruptly across regions, has been implemented earlier in~\cite{Fogli:2009rd,
Chakraborty:2009ej} where a two-flavor analysis of collective effects
is carried out.
Our survey involves a three-flavor treatment of collective effects, 
and is carried out over a different slice of the parameter space, 
i.e. we allow the $\nu_x$ energies to vary while fixing 
$L_{\bar{\nu}_e}=L_{{\nu}_e}$. 
The surveys in Ref.~\cite{Fogli:2009rd, Chakraborty:2009ej} chose to 
fix the $\nu_x$ energy and vary the relative fluxes / luminosities
of the flavors.
Our choice is motivated by the observation that most simulations tend to 
predict $L_{\nu_e}\approx L_{\bar{\nu}_e}$, whereas $\langle E_{\nu_x} \rangle$ 
turns out to be much less robustly predicted. 
We therefore believe that this survey provides useful information 
that is complementary to the one in~\cite{Fogli:2009rd,Chakraborty:2009ej}.

From the phase diagram, it is observed that for NH,
there are two phases with qualitatively different split patterns.
Phase A, which approximately corresponds to $L_{\nu_e} \gtrsim L_{\nu_x}$, 
shows no spectral swaps in $\nu$ or $\bar\nu$. On the other hand, in
the other phase C with $L_{\nu_e} \lesssim L_{\nu_x}$, 
one obtains a $\nu_e \leftrightarrow \nu_y$
swap and a $\bar\nu_e \leftrightarrow \bar\nu_y$ swap, both
at energies $\gtrsim 25$ MeV.
The situation is more complicated for IH. Here the phase A, which is
approximately the same as the phase A in NH, corresponds to a
$\nu_e \leftrightarrow \nu_y$ swap at energies $\gtrsim 8$ MeV,
and a $\bar\nu_e \leftrightarrow \bar\nu_y$ swap
at energies~$\gtrsim 4$~MeV.
The phase C however has different features depending on its sub-phases
${\rm C_1, C_2, C_3}$ and ${\rm C_4}$.
\begin{itemize}
\item ${\rm C_1}$ correponds to a large $L_{\nu_x}$ and 
large $\langle E_{\nu_x} \rangle$, and hence here the spectral
swaps are more pronounced.
Both $\nu_e \leftrightarrow \nu_y$ as well as 
$\bar\nu_e \leftrightarrow \bar\nu_y$ swaps appear at intermediate
energies $10$ MeV $\lesssim E \lesssim 25$ MeV. For $E \gtrsim 25$ MeV,
there are $\nu_e \leftrightarrow \nu_x$ and 
$\bar\nu_e \leftrightarrow \bar\nu_x$ swaps.
\item In ${\rm C_2}$, the energy $\langle E_{\nu_x} \rangle$ 
is lower as compared to ${\rm C_1}$. 
Then $\bar\nu_e$ and $\bar\nu_y$ spectra are very
similar at intermediate energies. 
As a result, the $\bar\nu_e \leftrightarrow \bar\nu_y$
swap at intermediate energies has low adiabaticity, and is
not effective.
\item In ${\rm C_3}$, the luminosity $L_{\nu_x}$ is lower as compared
to ${\rm C_1}$. Then the $\nu_e$ and $\nu_x$ spectra, as well as the
$\bar\nu_e$ and $\bar\nu_x$ spectra, are very similar at high energies. 
As a result, the $\nu_e \leftrightarrow \nu_x$ and
$\bar\nu_e \leftrightarrow \bar\nu_x$ swaps at high energies have 
low adiabaticity, and are not effective.
\item ${\rm C_4}$ corresponds to almost equal $\bar\nu_e$ 
and $\bar\nu_x$ number fluxes everywhere except at intermediate
energies near the peaks. As a result, the $\nu_e \leftrightarrow \nu_x$
and $\bar\nu_e \leftrightarrow \bar\nu_x$ swaps at high energies
are non-adiabatic and hence ineffective. 
The $\nu_e \leftrightarrow \nu_y$ and $\bar\nu_e \leftrightarrow \bar\nu_y$
swaps at intermediate energies are partially adiabatic. This case can be 
thought to be the combination of $\rm C_2$ and $\rm C_3$, with only the 
common features surviving.
\end{itemize}


In Fig.~\ref{spectra_after_collective}, we illustrate the spectral swaps
for  two benchmark points
 \begin{eqnarray}
 \hspace{-0.4cm}
{\rm A}:   \langle E_{\nu_x} \rangle &=& 18 \,\ \textrm{MeV},~ 
L_{\nu_x}/L_{\nu_e} = 0.7\;, \nonumber \\
{\rm C_1}:\langle E_{\nu_x} \rangle &=& 20 \,\ \textrm{MeV},~
L_{\nu_x}/L_{\nu_e} = 1.5\; . 
\label{eq:cases}
\end{eqnarray}  
The benchmark point A lies close to the predictions of the Garching
flux model \cite{garching} during accretion phase, while 
${\rm C_1}$ lies
close to the predictions during cooling phase.
Here we show the primary
fluxes of all $\nu$ as well as $\bar\nu$ species, and show how they
change after collective effects. 
The $\nu_e$ and ${\overline\nu}_e$ spectra after collective oscillations exhibit
the spectral features described before.  
 We point out  that for the $\rm C_1$ flux, in IH  
the oscillated $\nu_e$ spectrum shows a single  split, the  one at low-energy, corresponding to the swap with $\nu_y$, while the 
 high-energy split is canceled by the  further $\nu_e \leftrightarrow \nu_x$ swap. However, this high-energy split is still visible 
in the $\nu_x$ and $\nu_y$ spectra.  We will see in the following that this high-energy split would reappear
in the final  $\nu_e$ spectrum at Earth, since that is a superposition of the different $\nu$ spectra after collective 
oscillations.
The same observations apply for the $\bar{\nu}_e$ spectrum.
Moving from the case $\rm C_1$ to $\rm C_4$ in the phase $\rm C$ one may 
observe a  gradual degradation in the observability of the different spectral splits in final 
spectra (not shown).

\section{Fluxes after MSW conversions}
\label{sec:afterMSW}

\begin{table*}[!htpb]
\begin{center}
\caption{Survival probability $p$ for $\nu_e$ 
at low, intermediate and high energies,
for fluxes in phase A and C. 
Within phase C, the exceptions to the rule are denoted by the
brackets (..) for C$_3$ and C$_4$.
}
\begin{tabular}{|cc|ccc|ccc|}
\hline
& & \multicolumn{3}{c|}{Phase A ($L_{\nu_e} \gtrsim L_{\nu_x}$)}  & 
\multicolumn{3}{c|} {Phase C ($L_{\nu_e} \gtrsim L_{\nu_x}$)}  \\
& & $E<E_{\rm low} \quad$  & $E_{\rm low} < E < E_{\rm high} \quad$ & $E > E_{\rm high} $ &
 $E<E_{\rm low} \quad$  & $E_{\rm low} < E < E_{\rm high} \quad$ & $E > E_{\rm high} $ \\
\hline
\multirow{2}{*}{NH} & $\sin^2 \theta_{13} \gtrsim 10^{-3}$ &
0 & 0 & 0 &
0 & 0  & $\sin^2\theta_{12}$ \\
& $\sin^2 \theta_{13} \lesssim 10^{-5}$ &
$\sin^2\theta_{12}$ & $\sin^2\theta_{12}$ & $\sin^2\theta_{12}$ &
$\sin^2\theta_{12}$ & $\sin^2\theta_{12}$ & 0  \\
\hline
\multirow{2}{*}{IH} & $\sin^2 \theta_{13} \gtrsim 10^{-3}$ &
$\sin^2\theta_{12}$ & 0  & 0  &
$\sin^2\theta_{12}$ & 0  & $\cos^2\theta_{12}$ ($\sin^2\theta_{12}$)  \\
 & $\sin^2 \theta_{13} \lesssim 10^{-5}$ &
$\sin^2\theta_{12}$ & 0  & 0  &
$\sin^2\theta_{12}$ & 0  & $\cos^2\theta_{12}$ ($\sin^2\theta_{12}$)  \\
\hline
\end{tabular}
\label{fluxtable-p}
\end{center}
\end{table*}
\begin{table*}
\begin{center}
\caption{Survival probability $\bar{p}$ for $\bar\nu_e$ 
at low, intermediate and high energies,
for fluxes in phases A and C. 
Within phase C, the exceptions to the rule are denoted by:
brackets (..) for C$_3$ and C$_4$, square brackets [..] for ${\rm C}_2$ and ${\rm C}_4$.
}
\begin{tabular}{|cc|ccc|ccc|}
\hline
& & \multicolumn{3}{c|}{Phase A ($L_{\nu_e} \gtrsim L_{\nu_x}$)}
& \multicolumn{3}{c|} {Phase C ($L_{\nu_e} \gtrsim L_{\nu_x}$)}\\
& & $E<E_{\rm low} \quad$  & $E_{\rm low} < E < E_{\rm high} \quad$ & $E > E_{\rm high} $ &
 $E<E_{\rm low} \quad$  & $E_{\rm low} < E < E_{\rm high} \quad$ & $E > E_{\rm high} $ \\
\hline
\multirow{2}{*}{NH} & $\sin^2 \theta_{13} \gtrsim 10^{-3}$ &
$\cos^2\theta_{12}$ & $\cos^2\theta_{12}$ & $\cos^2\theta_{12}$ &
$\cos^2\theta_{12}$ & $\cos^2\theta_{12}$  & 0 \\
& $\sin^2 \theta_{13} \lesssim 10^{-5}$ &
$\cos^2\theta_{12}$ & $\cos^2\theta_{12}$ & $\cos^2\theta_{12}$ &
$\cos^2\theta_{12}$ & $\cos^2\theta_{12}$  & 0 \\
\hline
\multirow{2}{*}{IH} & $\sin^2 \theta_{13} \gtrsim 10^{-3}$ &
0 & $\cos^2\theta_{12}$ & $\cos^2\theta_{12}$ &
0 & $\cos^2\theta_{12}$ [0\,]  & $\sin^2\theta_{12}$ (0\,)  \\
 & $\sin^2 \theta_{13} \lesssim 10^{-5}$ &
$\cos^2\theta_{12}$ & 0 & 0 &
$\cos^2\theta_{12}$ & 0 [$\cos^2\theta_{12}$]  & $\sin^2\theta_{12}$ ($\cos^2\theta_{12}$)  \\
\hline
\end{tabular}
\label{fluxtable-pbar}
\end{center}
\end{table*}

\subsection{Survival probabilities}

After collective oscillation die out,  the primary
neutrino fluxes $F^0_{\nu}$ experience  
 spectral swaps and then further undergo the traditional MSW conversions
 in SN~\cite{Dighe,Fogli:2001pm}
while passing through the resonance regions $H$ and $L$, corresponding
 to flavor transitions in the two-neutrino sectors associated with 
 $(\Delta m^2_{\rm atm},\theta_{13})$
and $(\Delta m^2_{\rm sol},\theta_{12})$, respectively. 
After exiting the resonance regions, the neutrino mass eigenstates 
travel independently until they reach Earth, wherein they are 
detected as flavor eigenstates. The fluxes of $\nu_e$ and  
$\bar\nu_e$ arriving at Earth can be written as~\cite{Dighe} 
\begin{eqnarray}
F_{\nu_e} &=& p F^{0}_{\nu_e} + (1-p)F^{0}_{\nu_x} \,\ , \nonumber \\
F_{{\bar\nu}_e} &=& {\bar p} F^{0}_{{\bar\nu}_e} + (1-{\bar p})F^{0}_{{\nu}_x} \,\ ,
\end{eqnarray}
where $p$ and ${\bar p}$ are the $\nu_e$ and ${\bar\nu}_e$ 
survival probabilities, respectively.
The survival probabilities are determined by the adiabaticity of
the MSW $H$-resonance  and in general are sensitive to the neutrino energy 
and to the SN matter density  profile (see, e.g.,~\cite{Foglish}). 
This dependence vanishes for a large time-window in the limiting cases 
of adiabatic transitions ($\sin^2 \theta_{13} \gtrsim 10^{-3}$) 
and strongly non-adiabatic transitions ($\sin^2 \theta_{13} \lesssim 10^{-5}$)
(see, e.g.,~\cite{Lunardini:2003eh}).   
In the following,  we shall consider for simplicity only these two limiting   
cases.

In these two regions of the $\theta_{13}$ parameter space,
the expressions for $p$ and $\bar{p}$ can be written down in a
simple form, as shown in Tables~\ref{fluxtable-p} and \ref{fluxtable-pbar}.
The survival probabilities have been given separately for
the low-energy ($E<E_{\rm low}$), intermediate energy 
($E_{\rm low} < E < E_{\rm high}$) and high-energy ($E> E_{\rm high}$) regions.
Here $E_{\rm low}$ and $E_{\rm high}$, respectively, are the energies where 
the low-energy and high-energy spectral splits
are likely to be present.
For $\nu_e$, one has $E_{\rm low} \approx 10$ MeV and $E_{\rm high} \approx 25$ MeV.
For $\bar\nu_e$, one gets $E_{\rm low} \approx 5$ MeV and $E_{\rm high} \approx 25$ MeV.
The survival probabilities also depend on the primary fluxes,
and are given for the phases A ($L_{\nu_e} \gtrsim L_{\nu_x}$)
and C ($L_{\nu_e} \lesssim L_{\nu_x}$), covering the complete parameter
space. Some special subcases of C are given in the caption.
These are related to the cases 
 $\rm C_2$, $\rm C_3$, $\rm C_4$ shown in  Fig.~1, for which  we have seen that 
some of the splits can be suppressed by the presence of too close 
$\overline{\nu}_e$ and $\overline{\nu}_x$ spectra or by the low
${\nu}_x$ luminosity.

Note that the survival probability is never $1$. 
The vacuum mixing due to the angle $\theta_{12}$ always mixes the  
$\nu_e$ or $\bar\nu_e$ spectra with the other
flavors. Effects of the spectral swaps are invariably reduced by
a factor of $\sin^2 \theta_{12} \approx 0.3$ or
$\cos^2\theta_{12} \approx 0.7$.

\subsection{Earth matter effect} 
\label{sec:earth}
  
The neutrino survival probabilities at the Earth given in
Tables~\ref{fluxtable-p} and \ref{fluxtable-pbar} are calculated 
assuming that neutrinos escaping the star travel through vacuum before reaching
the detector.   
If the supernova is shadowed by the Earth for a detector, 
the neutrinos will travel a certain distance through the Earth, 
and will undergo Earth matter oscillations during this propagation.
Since the neutrinos arrive at the Earth as mass eigenstates,
the net effect of these oscillations can be written 
in terms of the conversion probabilities $P_{ie}=P(\nu_i \to \nu_e)$. 
Neglecting the effect due to small $\theta_{13}$, one can 
obtain the net survival probabilities by the substitution~\cite{Dighe}
\begin{equation} 
(\cos^2 \theta_{12}, \sin^2 \theta_{12}) \to (1-P_{2e}, P_{2e}) \,\  
\label{eme-nu}  
\end{equation}  
for neutrinos in Table~\ref{fluxtable-p}, and
\begin{equation}  
(\cos^2 \theta_{12}, \sin^2 \theta_{12}) \to (1 - \bar{P}_{2e}, 
\bar{P}_{2e}) \,\  
\label{eme-nubar}  
\end{equation}  
for antineutrinos in Table~\ref{fluxtable-pbar}.
Here
\begin{equation}  
P_{2e} \equiv P(\nu_2 \to \nu_e) \,\ , \quad \mbox{ and }
\bar{P}_{2e} \equiv P(\bar\nu_2 \to \bar\nu_e) \,\ 
\end{equation}  
while propagating through the Earth.
Analytical expressions for $P_{2e}$ and $\bar{P}_{2e}$ can be calculated
for the approximate two-density model of the Earth~\cite{Dighe:2003vm}.
When neutrinos traverse a distance $L$ through only the mantle of the Earth,
these quantities have a very simple form~\cite{Dighe,Lunardini:2001pb}:
\begin{eqnarray}  
P_{2e} & = & \sin^2\theta_{12} + \sin2\theta^m_{12} \times \label{P2e}\\
& & \sin(2\theta^m_{12}-2\theta_{12})  
\sin^2\left(  
\frac{\Delta m^2_{\rm sol} \sin2\theta_{12}}{4 E \,\sin2\theta^m_{12}}\,L  
\right)\,, 
\nonumber\\
\bar{P}_{2e} & = & \sin^2\theta_{12} + \sin2\bar\theta^m_{12} \times \label{Pbar2e}  \\
& & \sin(2\bar\theta^m_{12}-2\theta_{12})  
\sin^2\left(  
\frac{\Delta m^2_{\rm sol}\,\sin2\theta_{12}}{4 E \,\sin2\bar\theta^m_{12}}\,L  
\right)\,, \nonumber
\end{eqnarray}  
where $\theta^m_{12}$ and $\bar\theta^m_{12}$ are the effective
values of $\theta_{12}$ in Earth matter for neutrinos and antineutrinos,
respectively~\cite{Fogli:2001pm}.

The Earth crossing thus induces an oscillatory signature in   
the neutrino energy spectrum. 
However note that these oscillations are absent if the corresponding
survival probabilities vanish: they occur only when 
$p\neq0$ or ${\bar p} \neq 0$ in Tables~\ref{fluxtable-p} and 
\ref{fluxtable-pbar}.
Depending on the swap pattern, the Earth effect can then
appear in different energy regions in the spectra.
As we shall see in Sec.~\ref{sec:events}, the mere observation of Earth
matter effects is sometimes enough to distinguish between different
flux and mixing scenarios.
  
\section{Neutrino detection}  
\label{sec:detectors}

In this section we describe the main aspects and ingredients of 
our calculations  of supernova neutrino event rates.
The oscillated SN neutrino fluxes at Earth, $F_\nu$,  
must be convolved with the differential cross section   
$\sigma_e$ for electron or positron production, 
as well as with the energy resolution function $R_e$ of the detector, 
and the efficiency $\varepsilon$, in order to finally   
get observable event rates~\cite{FogliMega}: 
\begin{equation}  
N_e = F_\nu \otimes \sigma_e \otimes R_e \otimes \varepsilon\ .  
\label{Conv}  
\end{equation}  
{ To calculate the total number of events, we will assume the supernova distance $d=10$~kpc and integrate the event rates over $t=10$~s, assuming the fluxes to be constant over the entire duration. The fluxes are thus to be thought of as the time-averaged fluxes over 10~s. Alternately a more detailed modeling of the time-dependent flux is required, in which case a time-binned analysis may be performed.}

We will now describe the main 
characteristics of three types of detectors: water Cherenkov detectors, 
scintillation detectors, and liquid Argon Time Projection Chambers, 
that we have used to calculate the signals.    

\subsection{Water Cherenkov detectors}

In large water-Cherenkov detectors, the golden channel for supernova 
neutrino detection is the inverse beta decay of electron 
antineutrinos~\footnote{We will neglect the 
subleading neutrino interaction channels in the detectors, 
assuming that they can be separated at least on a statistical basis.}
\begin{equation}
{\bar\nu}_e + p \to n+ e^+ \,\ .
\end{equation}
For this process, we take the differential cross section 
from~\cite{Strumia:2003zx}.  
The total cross section grows approximatively as $E^2$.
We fold the differential cross sections for $e^{+}$ production with
a Gaussian energy resolution function of width $\Delta$. 
The value of $\Delta$ is predominantly determined by the
photocathode coverage of the detector. 
For our calculations we assume \cite{Tomas,Tomas:2003xn}
\begin{equation}
\Delta_{\rm{WC}}/\textrm{MeV} = 0.47\sqrt{E_e/\textrm{MeV}} \,\ ,
\end{equation}
where $E_e$ is the true positron energy. 
A galactic SN  is 
expected to produce  $\mathcal{O}(10^5)$ events in  
a Mt-class water Cherenkov detector with a fiducial 
volume of about 400~kt~\cite{Autiero:2007zj}.

\subsection{Scintillation detectors}

In liquid scintillators, the main channel for SN neutrino detection 
is the inverse beta decay of ${\bar\nu}_e$'s, the same as that in
water Cherenkov. However, here the positrons are detected through 
photons produced in the scintillation material. Since a larger
number of photons can be produced in a scintillation detector, 
these have typically a much better energy resolution than the 
water Cherenkov detectors. 
Indeed, the energy resolution
of a scintillation detector may be better by more than a factor of 6. 
The energy resolution of the scintillator detectors is determined 
by the number of photo-electrons produced per MeV, which for 
this type of detectors is expected to be given by
as good as~\cite{Wurm:2007cy,Oberauer}
\begin{equation}
\Delta_{\rm{SC}}/\textrm{MeV} = 0.075\sqrt{{E_e}/\textrm{MeV}} \,\ .
\label{sc-resolution}
\end{equation}
Since the Earth matter oscillations described in the
previous section may get smeared out by the finite energy resolution of the
detector, it is clear that the energy resolution plays a crucial role 
in the efficiency of detecting Earth effects.
For a fiducial mass of  50~kt, one expects 
 $\mathcal{O}(10^4)$ events
for a galactic SN~\cite{Autiero:2007zj} .
  
\subsection{Liquid Argon Time Projection Chambers}  
  
LAr TPC detectors would be particularly sensitive to SN electron 
neutrinos through their charged current interactions with Argon nuclei  
\begin{equation}  
 \nu_e + {}^{40}Ar \to {}^{40}K^{\ast} + e^- \,\ ,  
\end{equation}  
which proceed via the creation of an excited state of ${}^{40}K$ and its  
subsequent gamma decay. The Q-value for this 
inverse beta decay process is $1.505$~MeV. The cross-section 
for this charged current reaction is taken from \cite{Cocco:2004ac}.
Energy resolution in such detectors is expected to be very good.    
The energy resolution for leptons in liquid argon 
time projection chambers has been calculated by the 
ICARUS collaboration who report~\cite{GilBotella:2003sz}
\begin{equation}
\Delta_{\rm{LAr}}/\textrm{MeV} = 0.11\sqrt{{E_e}/\textrm{MeV}} + 
0.02\, E_e / {\rm MeV} \,\ .
\label{LAr-resolution}
\end{equation}
The fiducial volume for supernova neutrino detection is taken to be 
100~kt~\cite{GilBotella:2004bv}.   
With these standard inputs, one expects $\mathcal{O}(10^{4})$ 
events from the interactions considered above~\cite{Autiero:2007zj}, 
while the event rates produced by the other interaction channels 
are smaller by at least an order of magnitude.  
Due to the strong sensitivity to $\nu_e$, the liquid Argon technique 
would be complementary to the water Cherenkov and scintillation detectors, 
which are mostly sensitive to ${\bar\nu}_e$'s.

\section{Observables sensitive to neutrino flavor conversions}  
\label{sec:events}

In this Section we discuss about possible signatures of SN neutrinos flavor conversions
observable in the neutrino energy spectra in the three different types of detectors, presented
before. First we calculate the neutrino energy spectra at different detectors and we discuss
about the  observability of the spectral splits and of the Earth matter effect. Then, we also comment 
about two other possible observables sensitive to the neutrino mixing, namely the $\nu_e$ prompt neutronization
burst and the shock-wave effect on the SN neutrino signal.

\begin{figure*}[!t]
\centering
\begin{tabular}{cc}
\epsfig{file=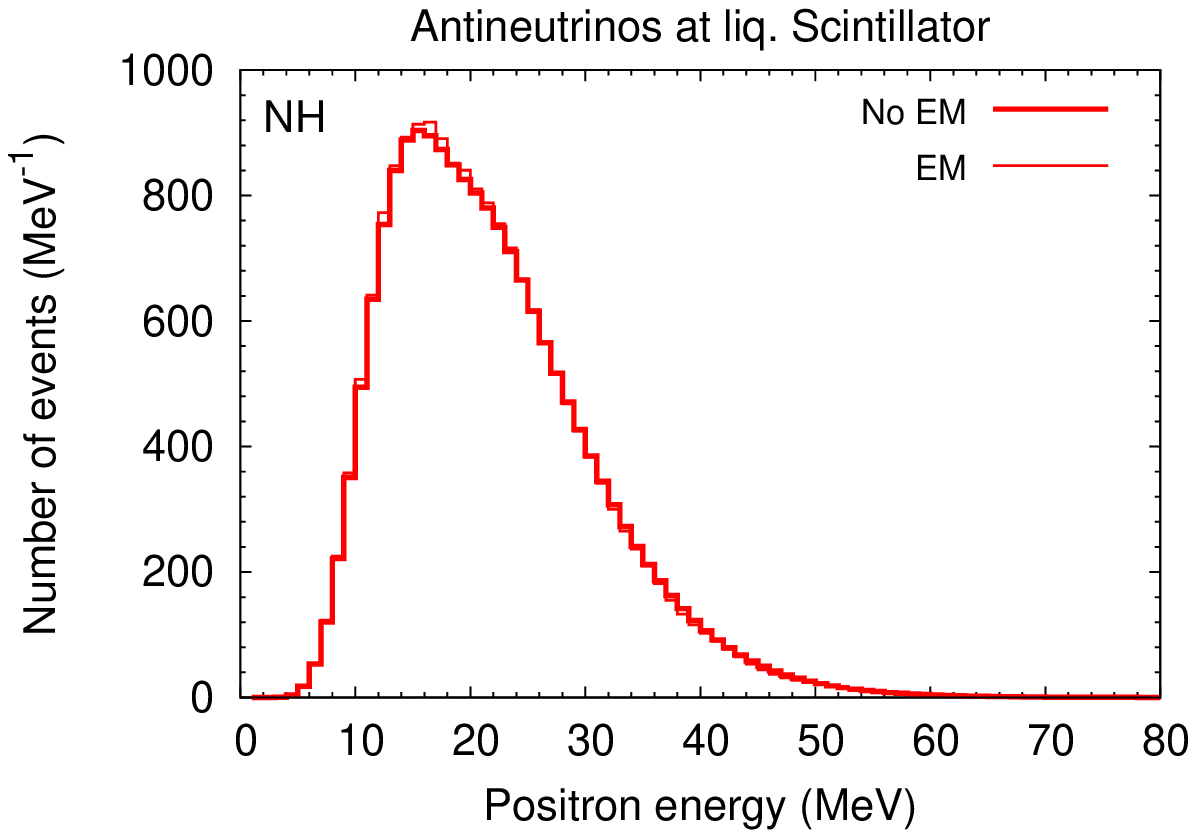,width=0.5\linewidth,clip=} &\epsfig{file=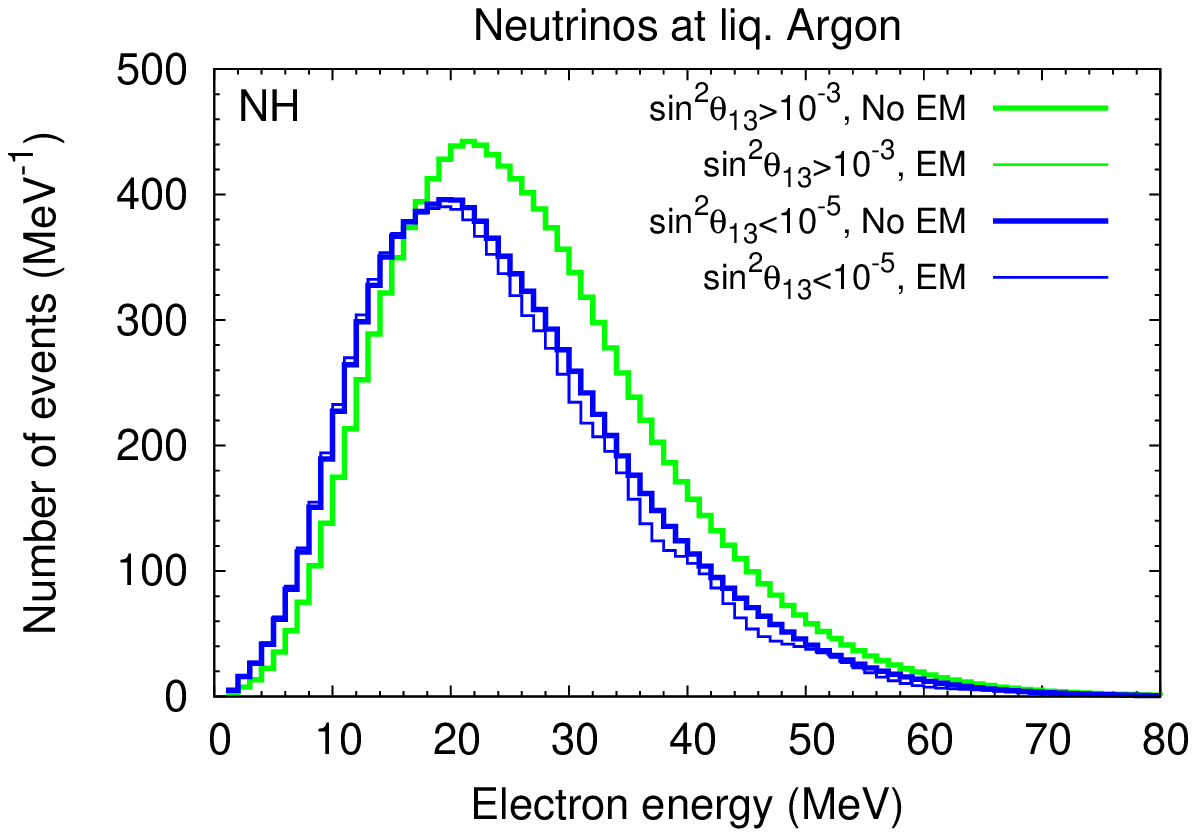,width=0.5\linewidth,clip=} \\
\epsfig{file=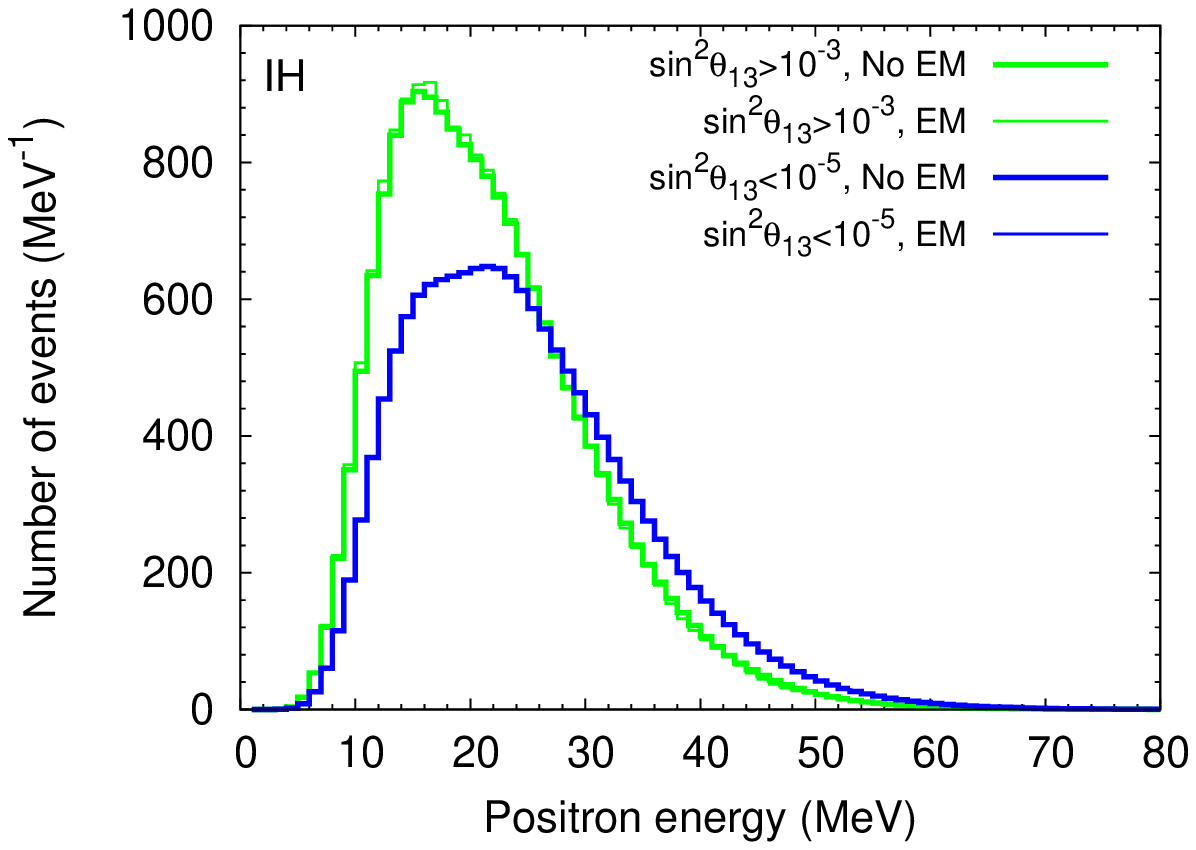,width=0.5\linewidth,clip=} &\epsfig{file=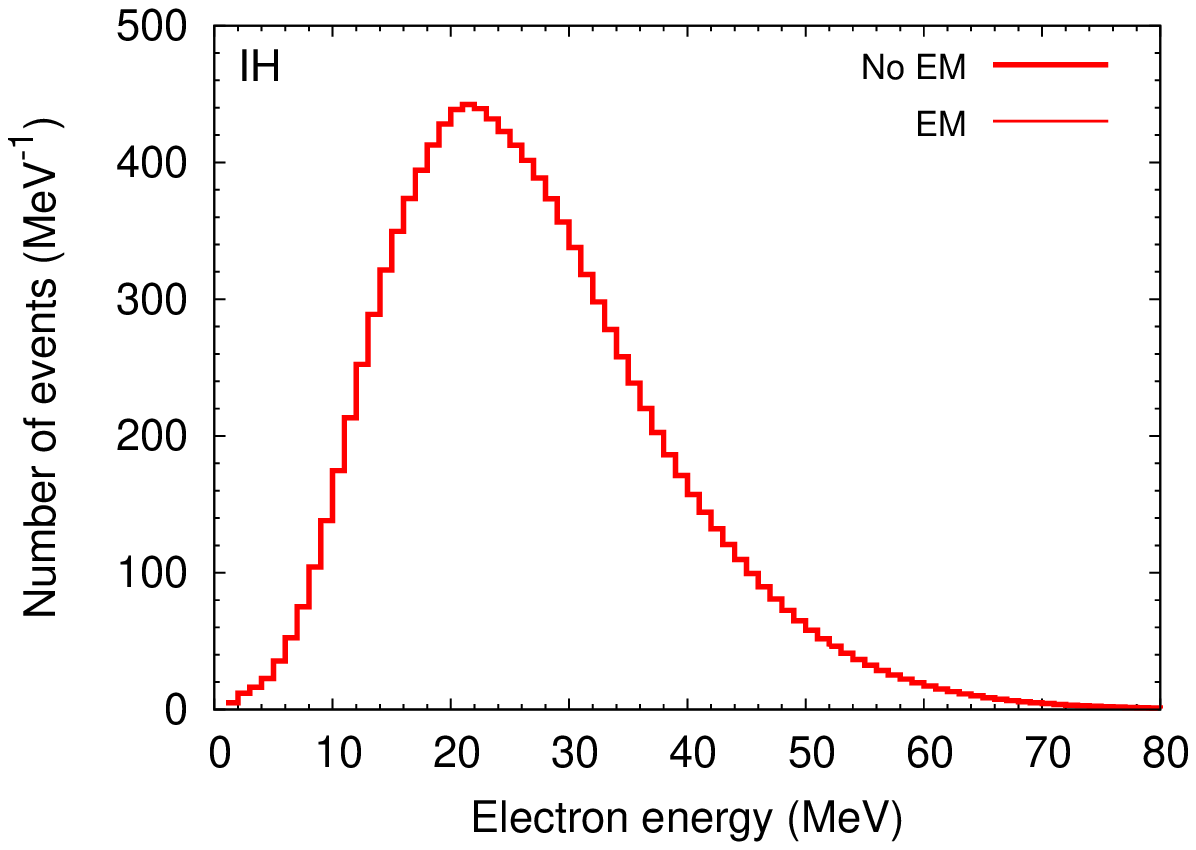,width=0.5\linewidth,clip=} \\
\end{tabular}
\caption{ For benchmark flux model A, 
$\bar\nu_e$ and  $\nu_e$ energy spectra at { $50$~kt} scintillator and { $100$~kt} LAr TPC detectors, in both the hierarchies NH (upper panels) and IH (lower panels), with and without oscillations
due to Earth matter effects. 
{ The spectra with Earth matter effects (EM) have been calculated for 
$L=8000$ km through the Earth, and have been denoted by thinner lines.}
The spectra at a water Cherenkov
detector can be obtained by smearing the energy of $\bar\nu_e$ at the
scintillator detector, and multiply the number of events  $\sim 10$ times.}
\label{spectra_at_detectors_A}
\end{figure*}

\begin{figure*}[!t]
\centering
\begin{tabular}{cc}
\epsfig{file=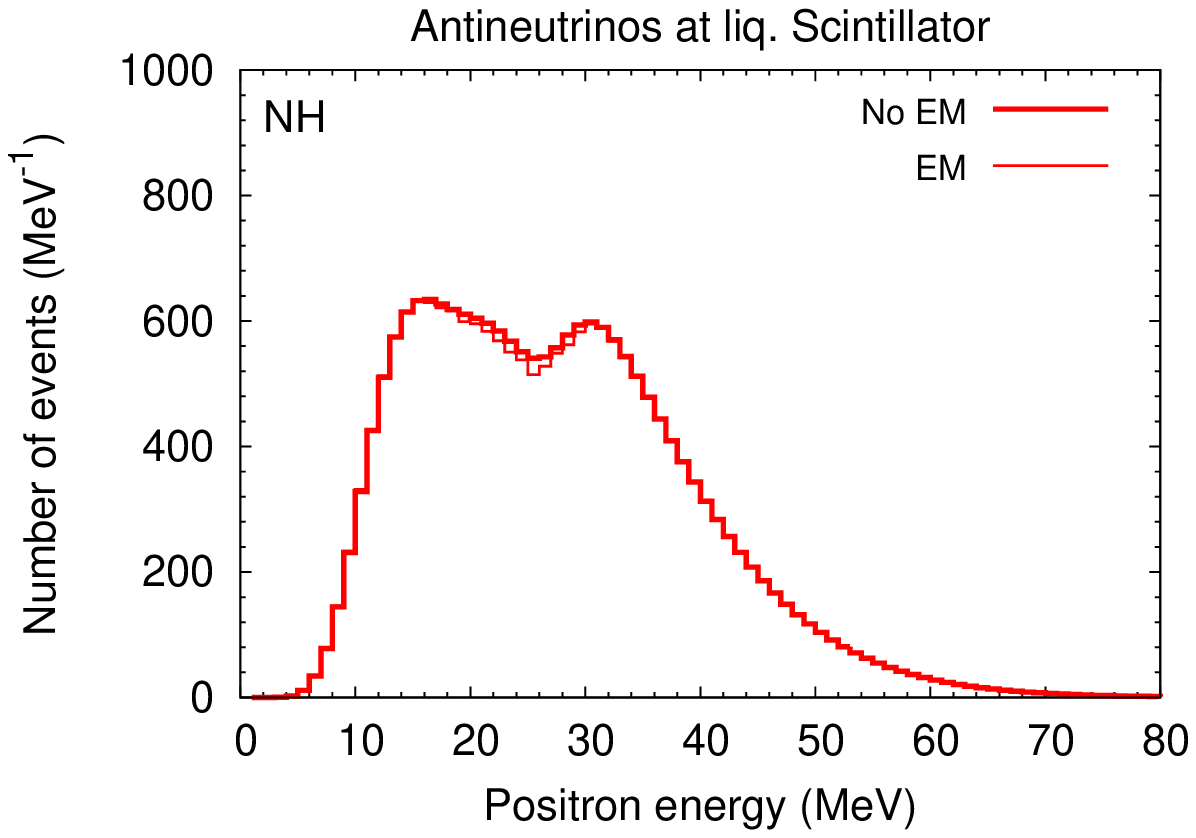,width=0.5\linewidth,clip=} &\epsfig{file=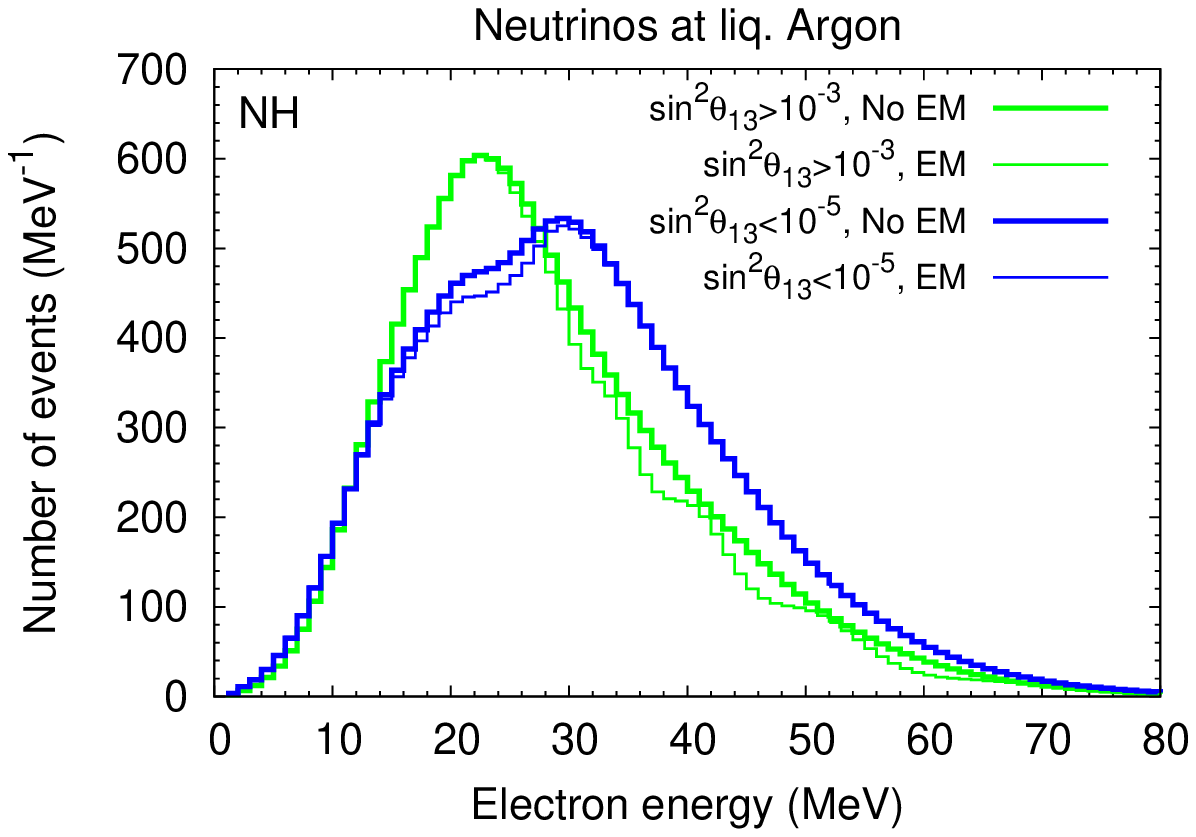,width=0.5\linewidth,clip=} \\
\epsfig{file=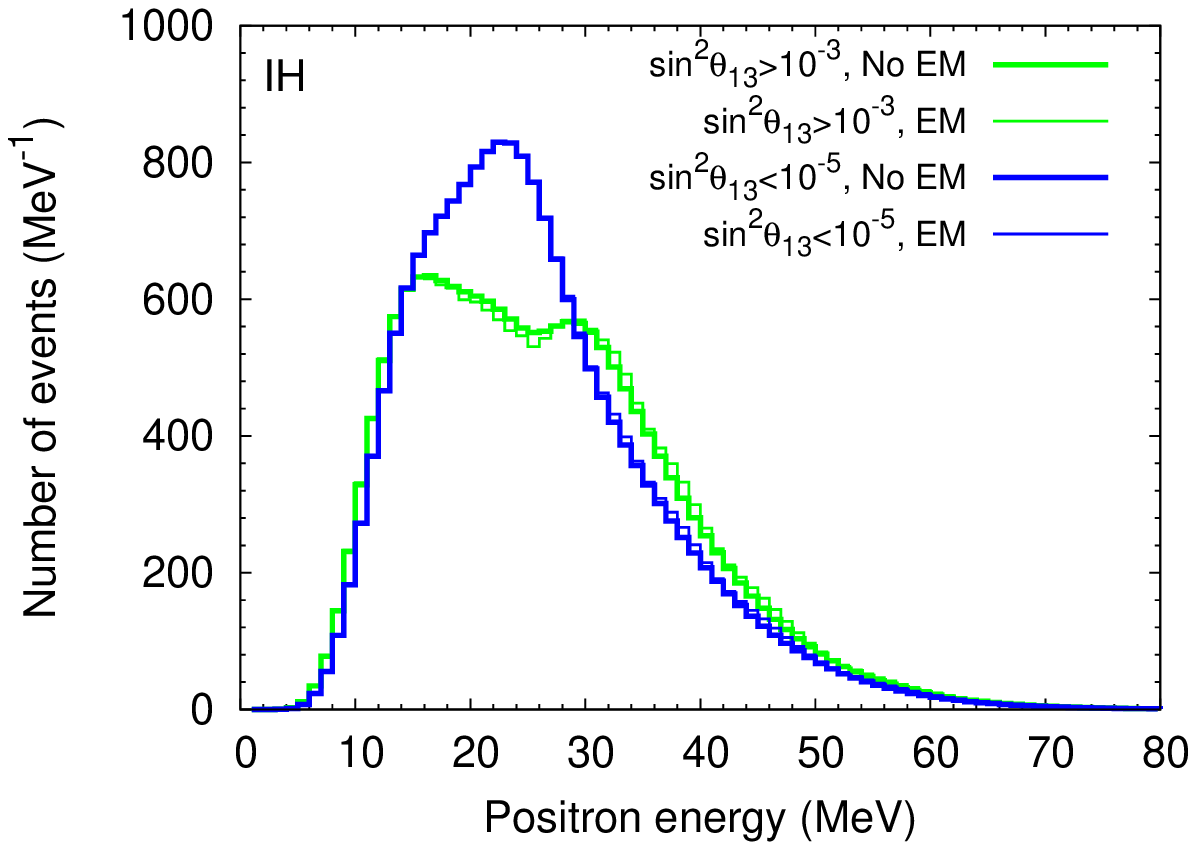,width=0.5\linewidth,clip=} &\epsfig{file=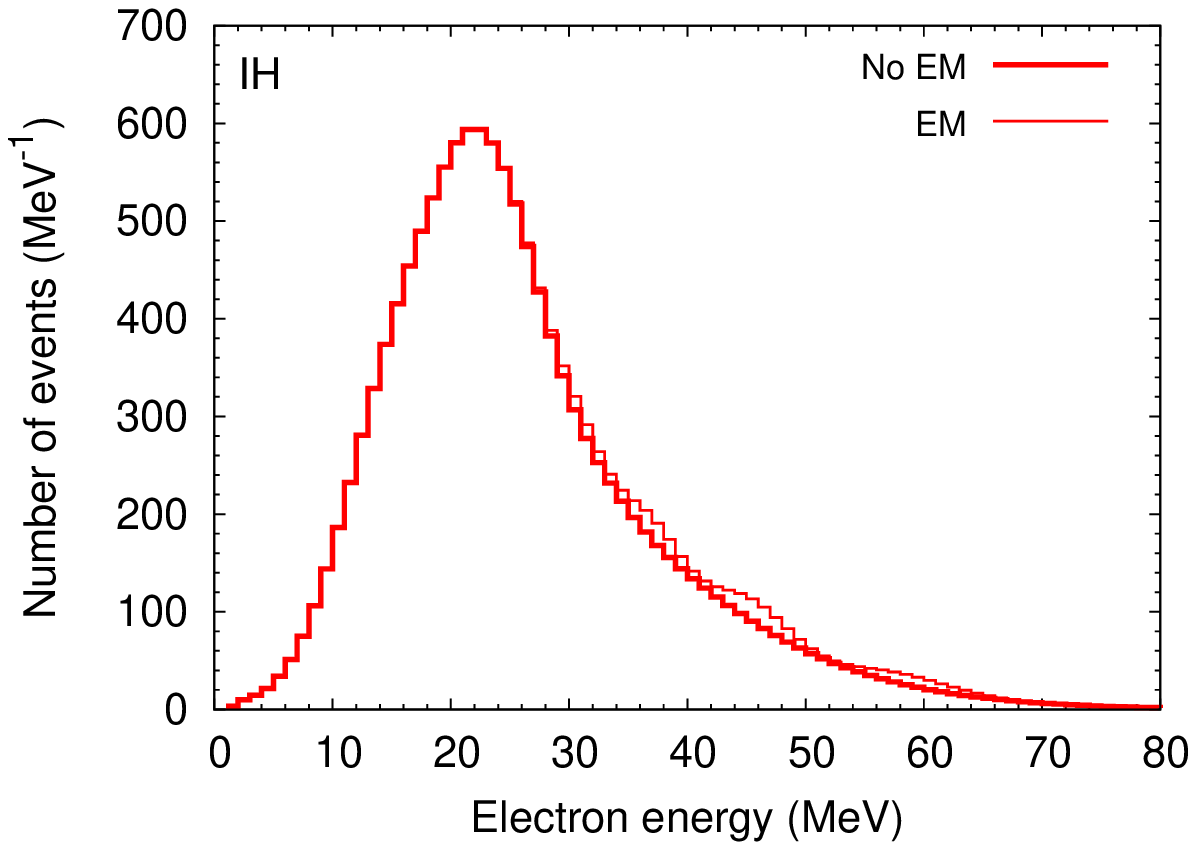,width=0.5\linewidth,clip=} \\
\end{tabular}
\caption{
The same as Fig.~\ref{spectra_at_detectors_A}, but for the
benchmark flux model ${\rm C_1}$.}
\label{spectra_at_detectors_C}
\end{figure*}

In order to get a realistic idea of what features of the neutrino
spectra can be observable at a neutrino detector, we show the 
events observable at different detectors for two of
our benchmark cases, A and ${\rm C_1}$, 
in Figures~\ref{spectra_at_detectors_A}
and ~\ref{spectra_at_detectors_C}, respectively. {  For the $\nu_e$ spectra, we show the events at a LAr TPC , while for the $\bar\nu_e$ spectra, we use the
 scintillation detector. The events spectra at the water Cherenkov
can be obtained by smearing the $\bar\nu_e$ spectra at 
the scintillation detector. We do not show them separately in this
figure. The event-rates will scale as $E_B/d^2$ and the 
volume of the detector, and one can read off the statistical errors, that go as $\sqrt{N}$, accordingly.}


We now analyze features in the detected neutrino spectra that
can tell us about the flux and mixing scenarios, and examine
the feasibility of detection of these features under different
conditions.

\subsection{Spectral splits}  

Spectral splits are produced at the boundaries of swaps, as shown in
Fig.~\ref{spectra_after_collective}. At a spectral split, the survival
probability $p$ or $\bar{p}$ jumps suddenly, so that a sharp jump 
in the spectrum is in principle observable if the primary spectra of
electron and non-electron flavors at the split energy are
different. The magnitude of the jump is reduced by the nonzero
value of $\theta_{12}$, and its sharpness is somewhat smeared out by
the energy resolution as well as possible non-adiabatic nature of the swap.

Figures~\ref{spectra_at_detectors_A} and \ref{spectra_at_detectors_C} 
bring out the following features of the spectral split:
\begin{itemize}
\item In the phase A, collective oscillations occur in IH, producing a spectral 
split in  the $\nu$'s and a complete swap of ${\bar\nu}$'s.
However, the split energy is relatively low (around $10$~MeV).
At this energy, the difference in $F_{\nu_e}$ and $F_{\nu_x}$ 
is small, and the charged current $\nu_e$-Ar cross section, 
being $\propto E^2$, is also small. As a result, this ``classic''
spectral split in phase A is practically unobservable,
as can be seen in Fig.~\ref{spectra_at_detectors_A}.
We realize that the peculiar interplay between collective and MSW effects 
can lead to quite different neutrino spectra for the two mass hierarchies and
for large and small values of $\theta_{13}$. However, lacking of a calibration
of the original SN neutrino fluxes, in this case it would be hard to make any strong statement 
about neutrino mixing just from the observation of the neutrino energy spectra.  

\item 
Observable spectra produced by $\nu$ with primary fluxes as in ${\rm C}_1$ 
are represented  in Fig.~\ref{spectra_at_detectors_C}. 
In the phase C, splits are possible in both, $\nu_e$ and $\bar\nu_e$
spectra at higher energies, around $E_{\rm{split}}=25$~MeV.
This is near the peak of the primary spectra, so the difference
between the electron and non-electron flavor spectra is more pronounced. 
This makes possible the detection of signatures related to spectral splits.
In particular for ${\overline\nu}_e$, in NH and in IH with 
$\sin^2 \theta_{13}\gtrsim 10^{-3}$,
the observable positron spectrum in a liquid scintillation detector 
(or in a water Cherenkov detector) is mostly due to 
$F^0_{\overline\nu_e}$ for $E < E_{\rm{split}}$
and $F^0_{\nu_x}$ at higher energies (see Table~II). 
This would produce a bimodal positron spectrum,
with two peaks corresponding to the two peaks of the initial 
antineutrino distributions.
Instead, at  $\sin^2 \theta_{13} \lesssim 10^{-5}$ { in IH},
the positron spectrum above  $E\simeq 10$~MeV
will be mostly produced   by  $F^0_{\overline\nu_x}$ (see Table~II), 
therefore no special spectral feature seems to be visible.

Concerning the electron spectrum produced by SN $\nu_e$ in a LAr TPC, 
in NH and  $\sin^2\theta_{13} \lesssim 10^{-5}$, 
we observe a bimodal distribution produced by
a superposition of    $F^0_{\nu_e}$ and  $F^0_{\nu_x}$ for 
$E\lesssim E_{\rm{split}}$ and by mostly $F^0_{\nu_x}$ at high energies 
(see Table~I), producing a broad   ``shoulder''. Instead, in NH and 
$\sin^2\theta_{13} \gtrsim 10^{-3}$ the electron
spectrum will be mostly produced by the  $F^0_{\nu_x}$. 
This is qualitatively similar to what happens also
in IH.

The presence of  bimodal distributions and  broad  
``shoulder'' features in the  spectra  produced by $\nu_e$ and $\bar\nu_e$   for the ${\rm C_1}$ primary flux, 
then hold 
the promise of being observable at the detectors,
in spite of the energy smearing.

\item In ${\rm C_2}$ the average energies of $\bar\nu_e$ and $\bar\nu_x$
are similar, as a result only one peak { would be} observed; the shoulder may 
not be discernible. Similarly in ${\rm C_3}$ and ${\rm C_4}$, since the
number fluxes of electron and non-electron flavors are close
to each other, direct observation of the spectral split is hard.
\end{itemize}
In summary, the spectral splits are directly identifiable only in the 
phase C, when the average energy and luminosity of
non-electron fluxes are sufficiently large.

\subsection{Earth matter effect }  

As discussed in Sec.~\ref{sec:earth}, the passage of  neutrinos 
through the Earth before reaching the detector can give rise to 
Earth matter effect oscillations in the spectra. 
For definiteness, in the following we will assume the neutrino path
length
in the Earth to be $L=8000$~km.
  
From Eqs.~(\ref{P2e}) and (\ref{Pbar2e}), these 
oscillations have frequencies that are functions only of the 
solar neutrino mixing parameters, and hence are well known. It is then
simply a question of identifying the oscillations in the final
spectra, if they exist. As can be seen in Figs.~\ref{spectra_at_detectors_A}
and \ref{spectra_at_detectors_C}, scintillation detectors and
LAr TPCs can allow us to detect these spectral modulations.
The following statements  can be made from  Table I and II and 
from  Figures 3 and 4:
\begin{itemize}
\item For the phase A, oscillations are expected in the 
$\nu_e$ spectrum for NH with $\sin^2 \theta_{13} \lesssim 10^{-5}$.
These oscillations should be clearly detectable in LAr TPC. 
 For NH with $\sin^2 \theta_{13} \gtrsim 10^{-3}$, as well as for IH,
there are no expected  Earth effects in the $\nu_e$ spectrum.
\item For the phase A, oscillations are expected in the
$\bar\nu_e$ spectrum for IH with $\sin^2 \theta_{13} \gtrsim 10^{-3}$ and in NH.
However in this case, the small flux differences between original 
antineutrino species leads to the presence of Earth induced oscillations
at high energies being barely visible, as can be seen from Fig.~3. 
The mixing scenarios IH with $\sin^2 \theta_{13} \lesssim 10^{-5}$
will not produce any Earth effects.
\item 
For ${\rm C_1}$, Earth effects in $\nu_e$ spectrum are expected in 
all mixing scenarios (See Table~I). In particular, 
for NH with $\sin^2~\theta_{13}~\lesssim~10^{-5}$, the Earth effects
are only at intermediate energies (10--25 MeV), while 
for the other mixing scenarios, they are prominent at high
energies ($E > 25$ MeV).
The sign of these effects is negative in NH and positive in IH.
In particular, as one can see from Fig.~4, that the Earth matter modulations 
are clearly visible in the electron spectrum at high energies.
This is due to the fact that in this case the spectral 
differences between  $F^0_{\nu_e}$ and $F^0_{\nu_x}$
are relatively large at high energies.
 
\item For the case ${\rm C_1}$, Earth effects in ${\overline\nu}_e$
are prominent at high energies if the hierarchy is IH.
For NH, the effects are expected at intermediate energies.
However, due to the smaller spectral 
differences between  $F^0_{\overline\nu_e}$ and $F^0_{\overline\nu_x}$
these features are difficult to be observed,
as can be seen from the Figure.
 
\item For C$_2$, C$_3$ and C$_4$ -- not shown in Figs.
~\ref{spectra_at_detectors_A} and  \ref{spectra_at_detectors_C} --
the Earth matter effects are in general less pronounced than
in C$_1$. This is a result of some of the swaps being absent
or only partially adiabatic.
\end{itemize}

\begin{figure}
\includegraphics[width=\linewidth]{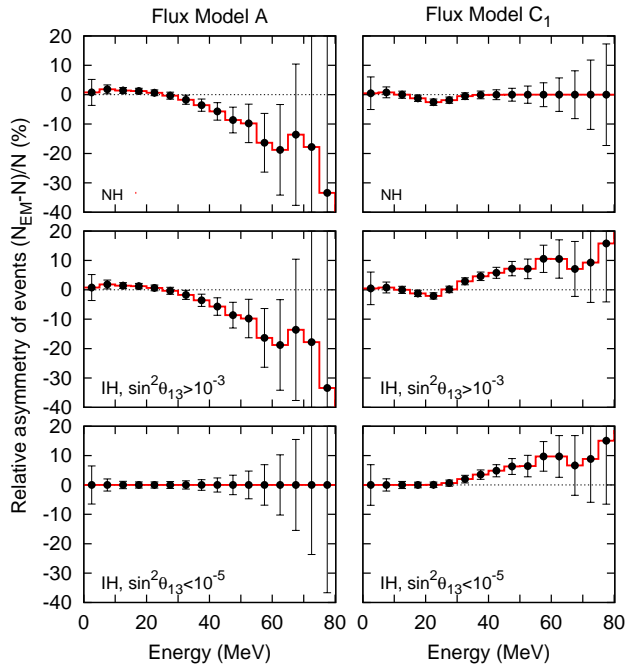}
\caption{ The left panel shows the relative asymmetry of $\bar\nu_e$ events with and 
without Earth matter effects at two $0.4$ Mt water Cherenkov detectors,
for the benchmark flux model A. For Earth effects, 
the distance travelled through the Earth is taken to be 8000 km. 
In the right panel we show the same ratio 
for the benchmark flux model $\rm C_1$.
}
\label{fig:ratios}
\end{figure}

The above observations are consistent with the prediction that 
the Earth effects would appear only at those energies where the 
survival probabilities $p$ or $\bar{p}$ in Tables
\ref{fluxtable-p} and \ref{fluxtable-pbar} are nonzero.

{ The Earth induced modulations can be identified at a single scintillation
or LAr TPC detector by the Fourier Transform technique~\cite{Dighe:2003jg,Dighe:2003vm}.} Clear observation of Earth effect oscillations in $\nu_e$
would then correspond to the possibilities
(i) primary fluxes A, and NH with~\mbox{$\sin^2\theta_{13}\lesssim 10^{-5}$}
(ii) primary fluxes C.
{ Observation of Earth effect oscillations in $\bar{\nu}_e$ flux at the 
shadowed detector is more likely for primary fluxes C, but may be possible for A as well.} The combination of the Earth effect signals in $\nu_e$ and $\bar{\nu}_e$ may then enable us to shortlist the possible combinations of the flux and mixing scenarios.
The Earth effects are visible in the intermediate or high energy range
depending on the flux and mixing scenario. 
If one is able to divide a spectrum into energy regions where
Earth effects are present and where they are not, further distinction
between flux and mixing scenarios is possible.
However practically this task looks very hard.
Also, note that non-detection of the Earth effects may be either due to
their non-existence or simply due to their being too small
to be detected. Therefore, in general the non-observation
of these oscillations would not give us any concrete information.

On the other hand, at a water Cherenkov detector whose energy
resolution may be insufficient for discerning the oscillations,
one may rely on comparison between signals at two
detectors. It is quite possible that we will have more than one
large water Cherenkov detectors within a couple of decades.
Moreover, IceCube can be used as an accurate luminosity
calibrator, and comparing the time dependence of luminosity 
in IceCube with that in another large water Cherenkov can 
help us identify Earth effects even without resolving the
oscillations~\cite{Dighe:2003be}.
Figure~\ref{fig:ratios} shows some scenarios where the
Earth effects may be observable without identification of 
oscillations, through the comparison between two detectors or 
the comparison between two different signals at the same detector.
A combination of two 0.4 Mt-class water Cherenkov detectors, one shadowed by the Earth
and the other un-shadowed~\cite{Mirizzi:2006xx}, 
has the capability to distinguish the mass hierarchies in the
phase A for $\sin^2\theta_{13} < 10^{-5}$, through an overall suppression 
of the event rate at the shadowed detector for NH~\cite{Dasgupta:2008my}.
In phase C, this combination can identify the hierarchies
through the enhancement of high-energy events at the shadowed detector. 
Note that one of the two detectors could in fact be 
IceCube, as we do not really need the energy dependence.

\subsection{Suppression of $\nu_e$ in the neutronization burst}

The primary signal during the early neutronization burst 
($t\lesssim 20$~ms) is pure $\nu_e$.
Since the model predictions for the energy and luminosity
of the burst are fairly robust~\cite{Kachelriess:2004ds},
the observation of the burst signal gives direct information
about the survival probability of $\nu_e$. This probability is
${\cal O}(\theta_{13}^2)$ in NH with $\sin^2 \theta_{13} \gtrsim 10^{-3}$,
and $\sin^2 \theta_{12}$ in all the other scenarios~\cite{Dighe}. 
Thus, the strong suppression of $\nu_e$ burst would be a smoking gun signal 
for the former scenario.
Note that since no $\bar\nu$ are involved, self-induced oscillations
are not developed and hence the collective effects do not give
rise to any flavor transformations during the neutronization burst.

However, in low-mass  O-Ne-Mg  supernovae ($M \simeq 8-10 M_{\odot}$), the MSW resonances may lie deep inside
the collective regions during the neutronization burst, when
the neutrino luminosity is even higher.
In such a situation, neutrinos of all energies undergo the
MSW resonances together, with the same adiabaticity~\cite{Wong:2002fa}.
As long as this adiabaticity is nontrivial, one gets the ``MSW-prepared
spectral splits'', two for normal hierarchy and one for inverted
hierarchy~\cite{Duan:2007sh, Duan:2008za, Dasgupta:2008cd,Cherry:2010yc}.
The positions of the splits can be
predicted from the primary spectra~\cite{Dasgupta:2008cd}.
The splits imply $\nu_e$ suppression that is stepwise in energy.
Such a signature may even be used to identify the O-Ne-Mg
supernova, in addition to identifying the hierarchy.
A LAr TPC with good timing and energy resolution would play a crucial role in this.
However, we leave the study of this particular case for a future work.

\subsection{Shock-wave effects}
\label{shock}

Observables like the number of events, average energy,
or the width of the spectrum may display dips or peaks 
for short time intervals, while the shock wave is
passing through the $H$-resonance.
The positions of the dips or peaks in the number of events at different 
neutrino energies would also allow one to trace the shock
propagation while the shock is in the mantle, around densities of
$\rho \sim 10^3$ g/cc~\cite{Tomas,FogliMega}. The interplay between collective 
oscillations and shock-wave signatures has been recently studied in~\cite{Gava:2009pj}.
This information, by itself or in combination with the corresponding
gravitational wave signal, will yield valuable information about 
the SN explosion, in addition to confirming 
$\sin^2\theta_{13} \gtrsim 10^{-3}$ and identifying NH (if the effects
are seen in the $\nu_e$ spectrum) or IH (if the effects are seen in
the $\bar\nu_e$ spectrum) as the actual hierarchy.
A caveat in this context is  the role of matter turbulences in the 
post-shock regions and their impact on neutrino flavor conversions. 
Presumably, these   would  at least partially erase the signatures
of the shock-waves in the neutrino signal~\cite{Fogli:2006xy, 
Choubey:2007ga, Friedland:2006ta, Kneller:2007kg,Kneller:2010sc}.

\begin{table*}
\begin{center}
\caption{Observable effects in the $\nu_e$ spectra, for fluxes of type
A and C. Earth effects at low energies ($E\lesssim 10$ MeV)
are almost impossible to detect and have not been considered.}
\begin{tabular}{|cc|ccc|ccc|}
\hline
& & \multicolumn{3}{c|}{Phase A ($L_{\nu_e} \gtrsim L_{\nu_x}$)}  & 
\multicolumn{3}{c|} {Phase C ($L_{\nu_e} \gtrsim L_{\nu_x}$) }  \\
& & ~~$\nu_e$ burst ~~ & ~~Earth effects ~~& ~~ Shock effects ~~&
~~$\nu_e$ burst ~~ & ~~Earth effects ~~& ~~ Shock effects ~~\\
\hline
\multirow{2}{*}{NH} & $\sin^2 \theta_{13} \gtrsim 10^{-3}$ &
Vanishes & Absent & Possible &
Vanishes & Only high $E$  & Possible \\
& $\sin^2 \theta_{13} \lesssim 10^{-5}$ &
Present & All $E$ & Absent &
Present & Only intermediate $E$ & Absent  \\
\hline
\multirow{2}{*}{IH} & $\sin^2 \theta_{13} \gtrsim 10^{-3}$ &
Present & Absent  & Absent  &
Present & Only high $E$  & Absent  \\
 & $\sin^2 \theta_{13} \lesssim 10^{-5}$ &
Present & Absent  & Absent  &
Present & Only high $E$  & Absent  \\
\hline
\end{tabular}
\label{tab:nue-effects}
\end{center}
\end{table*}

\begin{table*}
\begin{center}
\caption{Observable effects in the $\bar\nu_e$ spectra, for fluxes of type
A and C. The exceptions in region C are: 
with fluxes in regions C$_3$ and C$_4$ and IH.
Earth effects are absent for $\sin^2\theta_{13} \gtrsim 10^{-3}$,
while they are present at intermediate as well as high energies 
for $\sin^2\theta_{13} \lesssim 10^{-5}$.
Earth effects at low energies ($E\lesssim 10$ MeV)
are almost impossible to detect and have not been considered.}
\begin{tabular}{|cc|cc|cc|}
\hline
& & \multicolumn{2}{c|}{Phase A ($L_{\nu_e} \gtrsim L_{\nu_x}$) }  & 
\multicolumn{2}{c|} {Phase C ($L_{\nu_e} \gtrsim L_{\nu_x}$) }  \\
& & ~~Earth effects ~~& ~~ Shock effects ~~&
 ~~Earth effects ~~& ~~ Shock effects ~~\\
\hline
\multirow{2}{*}{NH} & $\sin^2 \theta_{13} \gtrsim 10^{-3}$ &
All $E$ & Absent &
Only intermediate $E$  & Absent \\
& $\sin^2 \theta_{13} \lesssim 10^{-5}$ &
All $E$ & Absent &
Only intermediate $E$  & Absent \\
\hline
\multirow{2}{*}{IH} & $\sin^2 \theta_{13} \gtrsim 10^{-3}$ &
Intermediate and high $E$  & Possible  &
Intermediate and high $E$  & Possible  \\
 & $\sin^2 \theta_{13} \lesssim 10^{-5}$ &
Absent  & Absent  &
Only high $E$  & Absent  \\
\hline
\end{tabular}
\label{tab:nuebar-effects}
\end{center}
\end{table*}

\section{Summary and conclusions}  
\label{sec:conclusions}

In this paper we have performed the first detailed study of the impact of 
collective and matter-induced flavor oscillations in the interpretation 
of the observable SN neutrino signal at a large water Cherenkov detector,
a scintillation detector, and 
a Liquid Argon Time Projection Chamber. 
We have analyzed the neutrino flavor evolution, including collective 
effects that give rise to spectral swaps and the MSW effects
that further mix the neutrino flavors. In particular, we have taken into 
account the possible qualitative change in the fluxes and oscillation physics 
during the SN neutrino emission.
We have calculated the neutrino flux at Earth, including possible
Earth matter effects  introduced if the neutrinos
pass through the Earth before reaching the detector.
We have also calculated the $\nu_e$ and $\bar\nu_e$ spectra
through the dominant channels at the above detectors, 
taking care of the detector efficiencies and energy resolutions.

The collective effects, which are dominant for $r \lsim 500$ km,
leave their imprints in the form of spectral swaps. 
The boundaries of spectral swaps are spectral splits, where
the $\nu_e$ or $\bar\nu_e$ spectra can have sharp jumps
at critical energy values.
Since the collective effects are essentially nonlinear, the
number and position of these splits depends on the primary
neutrino spectra.
To explore this, we have fixed the average energies for $\nu_e$ and 
$\bar\nu_e$ fluxes, assuming that their luminosities are almost equal 
(a result borne out by many SN neutrino simulations),
and we have scanned the parameter space in the average energy and luminosity of
the non-electron neutrinos.
It turns out that, depending on the number and nature of spectral
swaps, the parameter space may be divided into two main ``phases'':
the phase A ($L_{\nu_x} \lesssim L_{\nu_e}$)
that is typical of the fluxes during accretion,
and the phase C ($L_{\nu_x} \gtrsim L_{\nu_e}$)
that is typical of the fluxes during cooling.
Phase A shows no spectral swap in NH while in IH shows a spectral
swap at intermediate ($10$ MeV $\lesssim E \lesssim 25$ MeV) and high 
($E \gtrsim 25$ MeV) energies for both $\nu$ and $\bar\nu$.
Phase C, on the other hand, shows spectral swap 
in both $\nu$ as well as $\bar\nu$ at high energies even in NH.
In IH it in general results in two spectral swaps, 
a $e \leftrightarrow y$ swap at intermediate energies and
a $e \leftrightarrow x$ swap at high energies.

We have also  taken into account the MSW and Earth matter effects 
that determine the further
neutrino flavor conversions after the collective effects are over.
Armed with these results, we  have looked for distinctive signatures 
of neutrino mixing pattern in
the $\nu_e$ and $\bar\nu_e$ spectra at the detectors, and we have
examined the feasibility of their observation.
In particular, we have  considered 
(i) the sharp change in the spectrum that is the
signature of a spectral split (ii) the  Earth matter effects, 
(iii) the flavor conversion effects on the prompt $\nu_e$ neutronization burst,   and
(iv) the shock-wave effects in the neutrino as well as antineutrino
spectra.

The spectral split itself will be visible only if the corresponding
survival probability $p$ or $\bar{p}$ changes suddenly at
an energy $\gtrsim 10$ MeV. 
While the splits in phase A only occur at low energies,
the splits in phase C can occur at intermediate energies
and can be observable. These splits will be a clear signature
of the collective effects taking place deep inside the star, 
since no other known phenomenon can give rise to such a sharp
change in the neutrino spectrum.
Typically, the presence of spectral splits would produce bimodal 
observable energy spectral with two peaks corresponding to the 
ones of the initial $\nu_e$ and $\nu_x$ spectra. 
Our results for the other observables are summarized in 
Tables ~\ref{tab:nue-effects} and ~\ref{tab:nuebar-effects}.

From the tables, it can be observed that
\begin{itemize}
\item Whether the Earth effects are visible at
intermediate energies, high energies, or not at all, depends
on the neutrino mass hierarchy and the range of $\theta_{13}$.
A clear identification of the Earth effects is therefore 
crucial in extracting neutrino mixing information from the
observed spectra. (Note that since the threshold of the detectors 
will be $\sim 5$ MeV, we have ignored any effects at low energies.)
If the Earth effects are present in the intermediate energy range
and absent in the high energy range (or vice versa), it will be
a signature of a spectral split.
Thus, an indirect observation of a spectral split is also possible
through Earth matter effects.
\item The vanishing of $\nu_e$ burst and the observation of shock
wave effects is independent of the collective effects. These 
observables thus directy probe the neutrino mixing pattern:
mass hierarchy and $\theta_{13}$ range.
\end{itemize}
  
Concerning our study, we would to remind that while we have modeled 
the SN fluxes, and treated their oscillations 
in more detail than have been done in previous literature, 
the results shown here are 
for a simplified treatment of the SN neutrino problem which ignores possible 
trajectory-dependent effects, inhomogeneities, etc. 
All these effects would potentially introduce additional layers of 
complications in the simulation
of the supernova neutrino signal and their impact will  need 
careful dedicated studies. 
In this sense, we believe that  the predictions of possible signatures 
of supernova neutrino oscillations in large underground detectors need
further investigations.
This task is particularly timely now when different classes of 
large underground detectors are currently under study  
for low-energy neutrino astrophysics and long-baseline experiments. 
 In particular, from the perspective of supernova neutrinos 
it would be extremely useful
 to have as many different detection techniques as possible.
 
We would like to emphasize that the information obtained from 
the SN $\nu_e$ spectrum is as crucial as that obtained from the
$\bar\nu_e$ spectrum, and in some cases even more useful.
For example, the difference between the fluxes $F_{\nu_e}^0$ and 
$F_{\nu_x}^0$ is always much more than the difference between
the fluxes $F_{\bar\nu_e}^0$ and $F_{\nu_x}^0$.
Therefore, the features of flavor transformations like the
spectral splits and Earth effects are likely to be more prominent 
in the $\nu_e$ spectrum, if present.
Moreover, the shock-wave effects can be visible only in either
$\nu_e$ or $\bar\nu_e$ channel, depending on the mass hierarchy.
The information from the $\nu_e$ spectrum observed at a LAr TPC
would therefore be not just complementary to the one obtained
from a water Cherenkov or scintillation detector, but it will also 
probe features of the SN neutrino signal that are not accessible
to a $\bar\nu_e$ detector.

We wish to stress that the physics potential of large neutrino detectors 
proposed for low-energy neutrino astrophysics 
is immense, and can be exploited further for studies of SN neutrinos.
The accurate timing information obtained from water Cherenkov, 
scintillation and LAr TPC detectors should also 
allow us to track the SN $\nu$ light curve quite faithfully.   
A detailed study of the time evolution of the neutrino signal could 
offer new insights.  
A LAr TPC would also be the most efficient detector for observing the
$\nu_e$ neutronization burst. A comparison between the onset timing
of this signal with the onset time signals from $\bar\nu_e$ 
\cite{Pagliaroli:2009qy,Halzen:2009sm,Dasgupta:2009yj,Lund:2010kh} 
detectable in water (or ice) Cherenkov and scintillation detectors,
and gravitational wave signals
may shed more light on the SN dynamics and neutrino emission.

In conclusion, large future neutrino detectors would offer unprecedented 
opportunities to study supernova  neutrinos and determine fundamental 
neutrino properties through high-statistics studies of energy and time 
spectra.  With the complementary physics potential of future 
water Cherenkov, scintillation and liquid Argon detectors, 
they promise to advance our understanding of the physics of neutrinos, 
and of their flavor conversions during a stellar collapse.

\vspace{-0.5 cm}    
\begin{acknowledgments}  
A.D. was partly supported by the Max Planck - India Partner
Group in Neutrino Physics and Astrophysics. A.M. was supported by the 
German Science Foundation (DFG) within the Collaborative Research 
Center 676 ``Particles, Strings and the Early Universe".

\end{acknowledgments}


\end{document}